\documentclass{aastex701}
\usepackage{amsmath}
\usepackage{csquotes}
\begin{document}
\title{The Stellar Initial Mass Function Down To 0.16 M$_{\odot}$ Towards the Small Magellanic Cloud}

\correspondingauthor{Roger E. Cohen}
\email{rc1273@physics.rutgers.edu}

\author[0000-0002-2970-7435]{Roger E. Cohen}
\affiliation{Department of Physics and Astronomy, Rutgers the State University of New Jersey, 136 Frelinghuysen Rd., Piscataway, NJ, 08854, USA}
\email{rc1273@physics.rutgers.edu}
\affiliation{Eureka Scientific Inc., 2452 Delmer Street, Oakland, CA 94602, USA}

\author[0000-0002-5581-2896]{Mario Gennaro}
\affiliation{Space Telescope Science Institute, 3700 San Martin Drive, Baltimore, MD 21218, USA}
\affiliation{The William H. Miller {\sc III} Department of Physics \& Astronomy, Bloomberg Center for Physics and Astronomy, Johns Hopkins University, 3400 N. Charles Street, Baltimore, MD 21218, USA}
\email{gennaro@stsci.edu}

\author[0000-0001-6464-3257]{Matteo Correnti}
\affiliation{INAF Osservatorio Astronomico di Roma, Via Frascati 33, 00078, Monteporzio Catone, Rome, Italy}
\affiliation{ASI-Space Science Data Center, Via del Politecnico, I-00133, Rome, Italy}
\email{matteo.correnti@inaf.it}

\author[0000-0001-5538-2614]{Kristen B. W. McQuinn}
\affiliation{Department of Physics and Astronomy, Rutgers the State University of New Jersey, 136 Frelinghuysen Rd., Piscataway, NJ, 08854, USA}
\affiliation{Space Telescope Science Institute, 3700 San Martin Drive, Baltimore, MD 21218, USA}
\email{kmcquinn@stsci.edu}

\author[0000-0002-0572-8012]{Vedant Chandra}
\affiliation{Center for Astrophysics, Harvard \& Smithsonian, 60 Garden St, Cambridge, MA 02138, USA}
\email{vedant.chandra@cfa.harvard.edu}

\begin{abstract}

The presence (and nature) of variations in the stellar initial mass function (IMF) at substantially sub-solar masses and metallicities ($m$$<$0.5M$_{\odot}$, [M/H]$\lesssim$$-$1) remains poorly constrained.  Predictions from simulations vary widely, while observationally, resolved star studies of ultra-faint dwarf galaxies (UFDs) suffer from small sample sizes and background galaxy contamination due to low projected stellar densities.  As an alternative metal-poor target, we measure the IMF from resolved stars towards a carefully selected field in the Small Magellanic Cloud (SMC), leveraging a plethora of independent constraints on the target field stellar population including distributions of distance, 
age and metallicity.  We resolve $>$15,000 stars down to 0.16M$_{\odot}$ within a single pointing of NIRCam onboard JWST, using an observing strategy that minimizes contamination from point-source-like background galaxies.  We explore three different functional forms of the IMF, forward modeling observed color-magnitude diagrams (CMDs) and luminosity functions.  We find a best-fit single power law IMF slope of $\alpha$=$-$1.61$^{+0.03}_{-0.03}$, consistent with UFDs probed down to similar limiting masses.  Fitting a broken power law IMF, we find low- and high-mass slopes of $\alpha_{1}$=$-$1.44$^{+0.04}_{-0.04}$,  $\alpha_{2}$=$-$2.17$^{+0.11}_{-0.11}$ respectively, consistent with solar neighborhood values.  Assuming a lognormal IMF, we find a characteristic mass and lognormal width of $m_{c}$=0.12$^{+0.03}_{-0.03}$M$_{\odot}$, $\sigma$=0.61$^{+0.07}_{-0.06}$M$_{\odot}$, allowing for characteristic masses lower than local values as seen in some simulations as well as low-metallicity Galactic clusters.  Lastly, we quantify the impact of assumptions required in our analysis and discuss potential future improvements.

\end{abstract}

\section{Introduction}

The initial mass function (IMF) of stars with sub-solar masses, and its universality, remains one of the most pervasive unsolved problems in stellar astrophysics.  The relevance of potential variations in the low-mass IMF is difficult to overstate: Stars with subsolar masses contribute the majority of the present-day stellar mass budget, and the low-mass stellar IMF is a critical input needed to understand any observed stellar population.   
Assumptions on the IMF and its functional form are required, for example, to infer star formation histories (SFHs) and chemical evolution histories from either imaging of resolved stars or spectral energy distribution (SED) fitting of any stellar population (see reviews by \citealt{tolstoy09}, \citealt{conroy13}, \citealt{annibalitosi}), with potentially dramatic implications for cosmology and the early universe \citep[e.g.,][]{woodrum24}.  However, current evidence, both theoretical and observational, remains ambiguous regarding the impact of metallicity on the low-mass IMF.  

On one hand, the peak of the low-mass stellar IMF may be a function primarily of dust opacity and molecular hydrogen physics at the onset of star formation, with only a weak dependence on metallicity (e.g., simulations by \citealt{tanvir24}), 
with some models suggesting a universal IMF peak near $\sim$0.2M$_{\odot}$ independent of large-scale environment \citep{hennebelle18}.  On the other hand, some simulations predict that the IMF should depend on metallicity.  In particular, the cooling efficiency of protostellar gas clouds and the masses at which stars form can depend on metal content.  However, the detailed causes of such variation, which may also include the impact of redshift-dependent properties of the cosmic microwave background, may be difficult to disentangle (e.g., \citealt{adams96}, \citealt{larson98}, \citealt{bonnell06}, \citealt{bate25}, see \citealt{hennebelle24} for a review).  
In light of varying model predictions regarding the dependence of the low-mass stellar IMF on metallicity (or environment more generally), several IMF studies have targeted resolved stars in environments more metal-poor than the Milky Way.  Initially, Galactic globular clusters were targeted, but their present day mass function is difficult to tie to their IMF due to the various dynamical effects experienced over their lifetimes, including mass segregation, evaporation and in some cases stellar collisions, as well as uncertainties in gas expulsion timescales \citep[][also see the review by \citealt{bastian10}]{bm03,bk07,paust10}.  Young clusters within the Milky Way that are \textit{relatively} metal-poor have also been targeted, but are suboptimal because of their higher metallicities ([M/H]$>$$-$1), sensitivity to assumed distance, and reliance on pre-main-sequence evolutionary tracks and difficulties in precisely mapping dust attenuation 
in the face of high internal extinction 
\citep[e.g.,][]{dario12,yasui23,yasui24,andersen25}.  

Given such drawbacks involved in using targets within the Milky Way to constrain the metallicity dependence of the low-mass IMF, 
an attractive alternative set of targets 
has included the SMC and several nearby ultrafaint dwarf (UFD) galaxies (due in part to their much longer dynamical timescales, 
e.g., \citealt{spitzer87}; also see \citealt{geha13}).  
Initially, studies of the subsolar-mass IMF in these nearby metal-poor dwarfs 
quantified their results by measuring a single power-law slope (denoted $\alpha$) to compare against the canonical \citet{salpeter} value of $\alpha$=$-$2.35.\footnote{We adopt the definition of the power-law slope $\alpha$ advocated by, e.g., \citealt{hopkins18} and \citealt{jerabkova25}, in which the negative sign is included in the value of $\alpha$ according to the definition d$N$/d$m$ $\propto$ $m^{\alpha}$, with $m$ in units of M$_{\odot}$.}  
However, it has been known since the work of \citet{scalo86} that the observed luminosity function in the Milky Way is incompatible with a single power law when analyzed well into the sub-solar mass regime.  The low mass stellar IMF in the disk and bulge of the Milky Way has more recently been characterized as either a broken power law \citep[][also see \citealt{ktg93,kroupa13}]{kroupa01} or lognormal function \citep{chabrier03,chabrier05}, which are essentially indistinguishable barring exquisite star count data down to $\lesssim$0.2M$_{\odot}$ \citep[see e.g., Figs. 1 and 3 of][]{offner16} \footnote{Alternative forms for the low-mass IMF aside from a broken power law or lognormal have been presented as well \citep[e.g.,][]{demarchi05,parravano11,maschberger13}.}.  Therefore, observational studies lacking the depth and mass range to fully characterize the \enquote{turnover} in the subsolar-mass IMF used single power law fits as the best available option to assess its universality \citep[e.g.,][]{bochanski10,sollima19,li23,wang25}. 

For example, a power-law IMF slope of $-$1.8 was reported for the Ursa Minor dwarf spheroidal galaxy by \citet{wyse02}.  Prompted by this result, IMF measurements were subsequently obtained for several UFDs as well as the SMC, generally finding slopes flatter (i.e., more bottom-light) than the Salpeter value at varying levels of statistical significance \citep{geha13,kalirai13,gennaro18a,gennaro18b,safarzadeh22,filion22,filion24}.  However, simulations demonstrated that this could be a natural consequence of observing an underlying IMF that is truly lognormal (or multipartite power law), when sampled over a limited mass range and fit with a single power law \citep{elbadry17}.  Furthermore, the ensemble of aforementioned studies, mostly targeting UFDs, often sampled different stellar mass ranges on the main sequence.  Differences in sampled mass range alone would yield different slopes when fitting single power laws to an underlying lognormal IMF (by $\sim$0.2 according to \citealt{yan24}; also see, e.g., Fig.~1 of \citealt{hennebelle24}).  Indeed, among existing IMF studies of metal-poor UFDs, those with sufficient observational depth to fit broken power law and/or lognormal IMFs have found that they do not differ from solar neighborhood values at a statistically significant level \citep{gennaro18a,gennaro18b,filion22,filion24}, and \citet{gennaro18b} found that their target UFD was similarly well fit down to $\sim$0.23M$_{\odot}$ by all three parameterizations of the IMF that were investigated (single power law, broken power law and lognormal).  

These previous investigations of the low-mass IMF, mostly targeting UFDs (with the exceptions of \citealt{wyse02} and \citealt{kalirai13}), have all been limited by some combination of photometric depth, number statistics, and/or background galaxy contamination.  To circumvent all of these issues, we exploit the throughput, spatial resolution and simultaneous near- and mid-infrared imaging capability of NIRCam onboard JWST to target a carefully selected field in the SMC while minimizing background galaxy contamination.  The SMC is a particularly attractive target for IMF studies using resolved stars for two reasons: First, its combination of proximity (D$_{\odot}$$\approx$62 kpc; \citealt{degrijsbono15}) and projected stellar density provide an ideal compromise between maximizing stellar number statistics while avoiding severe crowding.  For example, we detect $>$10$^{4}$ sources in our target field down to $\sim$0.16M$_{\odot}$ in a single NIRCam pointing, at a projected density of only $\sim$1/arcsec$^{2}$.  
  
Second, excluding the more crowded inner few kpc of the SMC, the mean metallicity drops to well below 10\% solar in its outer regions where ancient stellar populations dominate, providing a broad metallicity baseline for comparison against the Milky Way for constraining any metallicity dependence of the low-mass stellar IMF.  Additionally, an important ancillary benefit to targeting the SMC is the variety of external constraints that aid in IMF analyses, including multi-band ground-based imaging \citep[e.g.][]{nidever21}, spatially resolved analyses of extinction \citep[e.g.,][]{haschke11,gorski20,skowron21,oden25}, line-of-sight distance distribution \citep{sub17,tatton21,elyoussoufi21,omkumar21}, age-metallicity relation \citep{carrera08,piatti12}, and star formation history \citep[SFH; e.g.,][]{noel09,cignoni13,rubele18,cohen24b}.   

The remainder of this study is organized as follows: In Sect.~\ref{photsect}, we describe the selection of our target field within the SMC as well as our observations and PSF photometry.  In Sect.~\ref{anasect} we describe our method for fitting different functional forms of the IMF to our observed photometric catalog, and the resultant fits are presented in Sect.~\ref{resultsect}.  In Sect.~\ref{discusssect} we place our results in context and describe limitations and potential future improvements, and in Sect.~\ref{conclusion}, we summarize our findings.

\section{Data \label{photsect}}
\subsection{Target Field Selection \label{targetsect}}

Within the SMC, our target field location was selected based on three criteria, assessed using the wealth of ancillary information available.  First, we sought out low-metallicity fields with extant constraints on both their SFH from resolved imaging (\citealt{noel09}; also see \citealt{rubele18}) as well as their chemical evolution history 
(\citealt{carrera08}; also see \citealt{povick23}).  Second, among the candidate fields studied by \citet{carrera08} and \citet{noel09}, we required sufficient projected stellar density to observe N$>$15,000 sources down to well below $\sim$0.2M$_{\odot}$ within a single NIRCam field of view, effectively eliminating candidate fields beyond a few kpc from the SMC.  Third, we require fields as devoid as possible of young ($<$300 Myr) stars.  Below such ages (approximately the pre-main sequence lifetime of a $\sim$0.2M$_{\odot}$ star; \citealt{tognelli11}; also see \citealt{chabrier00}), our IMF analysis could be impacted by systematics due to changes in the mass-luminosity relation at young stellar ages.  With these parameters in mind, projected source densities and the fraction of young ($\lesssim$300 Myr) stars were assessed using publicly available photometric catalogs from SMASH DR2 \citep{nidever21} to select an optimal target field.  

\begin{figure}
\gridline{\fig{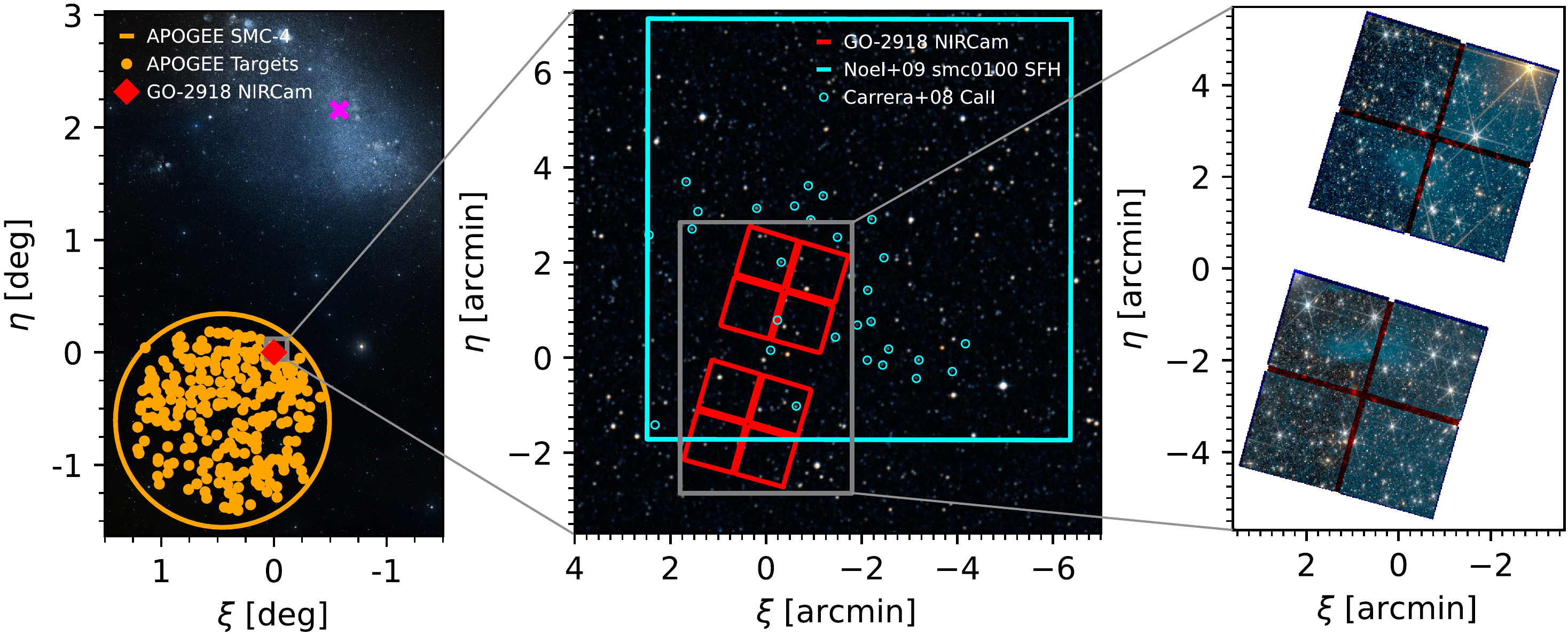}{0.99\textwidth}{}}
\caption{\textbf{Left: }The location of our NIRCam pointing (red diamond), overplotted on a DSS-2 image.  Axes are labeled with tangent plane coordinates relative to our NIRCam field center in all panels.  The APOGEE SMC-4 field is indicated by an orange circle, with individual SMC member targets shown using orange points.  The optical center of the SMC \citep{crowl01,subsub12} is indicated by a magenta cross.  \textbf{Center: }The footprints of the individual NIRCam SW detectors, again overplotted on a DSS-2 image.  Spectroscopic CaII targets from \citet{carrera08} are indicated with cyan circles, and the cyan box indicates the smc0100 optical imaging field used to infer the SFH presented in \citet{noel09} that we assume for our IMF analysis (see Sect.~\ref{inputsect}).  \textbf{Right: }A color image of our NIRCam field, where blue and red represent F150W and F322W2 respectively and green represents the mean of the two.
\label{fovfig}}
\end{figure}

The location of our selected target field is illustrated in Fig.~\ref{fovfig} in tangent plane coordinates relative to the target field center, located at (RA$_{\rm J2000}$, Dec$_{\rm J2000}$) = (15.15$^{\circ}$,-74.995$^{\circ}$).  Lying $\sim$2$^{\circ}$ south of the SMC main body, our NIRCam pointing is cospatial with the smc0100 field from \citet{carrera08} and \citet{noel09}, providing SFHs and chemical evolution histories, as well as one of the fields (SMC-4) from the Apache Point Galactic Evolution Experiment (APOGEE) H-band spectroscopic survey, providing an additional spectroscopic metallicity distribution {(spectroscopic metallicity distributions are discussed in Appendix \ref{testsect})}.   

\subsection{Observations and Photometry}

The target field was observed on 26 June 2024 (JWST-GO-2918, PI: Cohen) using the \texttt{MEDIUM8} readout pattern.  A 16-point subpixel dither pattern was used to optimize PSF sampling, with 8 groups per integration (for cosmic ray mitigation) and 1 integration per exposure, for a total exposure time of 13,399s (65\% of the total required time including overheads).  NIRCam imaging was obtained simultaneously in the F150W filter of the short wavelength (SW) channel and the F322W2 filter of the long wavelength (LW) channel.  While the F150W2 filter would have provided higher throughput in the SW channel, our filter pair was chosen to optimize rejection of background galaxies by maximizing their color separation from the stellar evolutionary sequences of true SMC sources (see Sect.~\ref{bgsect}).  

Preprocessing of all individual \texttt{uncal} science images was done with the JWST calibration pipeline version 1.17.1.dev727+ge6598024 (\texttt{pmap} version jwst\_1361.pmap, CRDS context 12.1.4).  
We made three modifications from the default pipeline settings: First, we set \texttt{suppress\_one\_group}=False when running Stage 1 of the pipeline, allowing the use of \enquote{Frame 0}, the first frame in each group of averaged frames, to extend our dynamic range brightward by $>$2 mag.  Second, we used \texttt{clean\_flicker\_noise} to mitigate 1/f noise, although this ultimately had no discernible impact on the quality of our photometry.   
Third, deep distortion-corrected \texttt{i2d} images, built by combining individual Stage 2 \texttt{cal} science images and required as a reference for PSF photometry, were generated on a per-detector basis in each filter.  We found that when using a single full-frame \texttt{i2d} reference image per filter, detector-dependent photometric offsets were seen in the resulting photometric catalogs.  These offsets were substantially (32\%) reduced (but not eliminated; see below) when photometry was performed on a per-detector basis before combining the results into a master catalog. 
  
Additional preprocessing and point-spread-function fitting (PSF) photometry was performed with the publicly-available DOLPHOT code \citep{dolphot,dolphot2}\footnote{\url{http://americano.dolphinsim.com/dolphot/}}.  DOLPHOT includes model PSFs tailored to each filter of each camera, and the application of DOLPHOT to generate photometric catalogs using NIRCam imaging is discussed in detail by \citet[][also see \citealt{savino24}]{weisz24}\footnote{Additional information and tutorials on the use of DOLPHOT with NIRCam imaging can be found at \url{https://dolphot-jwst.readthedocs.io/en/latest/}}.  Before performing PSF photometry, additional per-image preprocessing steps are required by DOLPHOT, including identifying bad and saturated pixels, applying a pixel area map and calculating a preliminary sky background frame.  

When performing PSF photometry using DOLPHOT, we use the parameters recommended by \citet{weisz24}, obtained by testing over a variety of crowding conditions.  As in other cases of imaging in two NIRCam filters with substantially differing PSF sizes (i.e., one SW and one LW-channel filter), our photometric catalogs were optimized by performing two separate photometry runs using the available warmstart mode within DOLPHOT \citep[e.g.,][also see \citealt{correnti25}]{mcquinn24,newman24}.  Specifically, an initial PSF photometry run was performed on the F150W images alone, followed by a second run on the full set of images (including F322W2) in warmstart mode, fixing the positions of the sources to those obtained from the F150W-only run.  The output catalogs from each run were then matched to obtain a master catalog with PSF photometry in both filters for each chip (there were no issues of potential mismatches since the positions in the second run were fixed based on the first run).  

The photometric catalogs output by DOLPHOT contain magnitudes in the Sirius-based Vegamag photometric system \citep[e.g.,][]{rieke22} which we use throughout this study unless otherwise stated.  To retain only well-measured stellar sources in our photometric catalogs, DOLPHOT provides several photometric quality parameters which we use to eliminate spurious, non-stellar and/or poorly-measured detections: 

\begin{enumerate}
\item Object type = 1
\item Per-filter reported signal-to-noise SNR$\geq$4
\item Per-filter photometric quality flag $\leq$2.  This flag is used to eliminate sources with too many bad or saturated pixels to provide reliable PSF photometry.  
\item Per-filter $\lvert$\texttt{sharp}$\rvert$$\leq$0.1.  The \texttt{sharp} parameter quantifies the central concentration of the flux distribution relative to the model PSF, and is positive for more compact sources (e.g., cosmic rays) and negative for less compact sources (e.g., blends or unresolved galaxies).  
\item \texttt{crowd}$_{\rm F150W}$ $\leq$0.15, \texttt{crowd}$_{\rm F322W2}$ $\leq$0.3.  The \texttt{crowd} parameter quantifies how much brighter a source would have been (in magnitudes) had its neighbors not been iteratively subtracted when measuring its flux.  
\end{enumerate}

\begin{figure}
\gridline{\fig{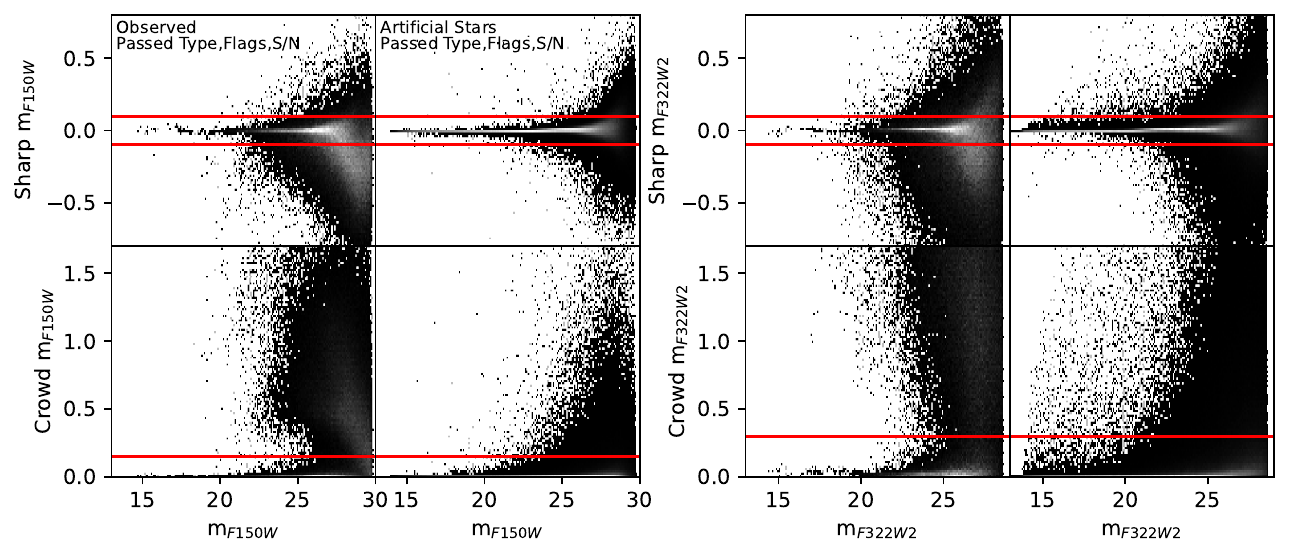}{0.99\textwidth}{}}
\caption{\textbf{Left: }The photometric quality parameters \texttt{sharp} (top row) and \texttt{crowd} (bottom row) shown as a function of m$_{\rm F150W}$, magnitude in the F150W filter.  Our chosen \texttt{sharp} and \texttt{crowd} cuts, shown using horizontal red lines, were set by comparing the distribution versus m$_{\rm F150W}$ seen in our catalog (left column) against the distribution seen for artificial stars (right column) to minimize the occurrence of spurious detections not present in the artificial star sample.  Only sources passing the other photometric quality cuts (object type, photometric quality flag, signal-to-noise) are shown.  \textbf{Right: }Same, but versus  m$_{\rm F322W2}$. 
\label{qparam_fig}}
\end{figure}

These quality cuts were guided by previous DOLPHOT PSF photometry of NIRCam imaging over a variety of crowding conditions \citep[e.g.,][]{mcquinn24wlm,mcquinn24,newman24,weisz24,cohen25}.  While the goal is to retain as many real stellar sources as possible, thorough rejection of spurious and non-stellar sources is critical in the present case since our analysis techniques and best-fit metrics rely on using synthetic photometry to replicate the observed color-magnitude distribution (see Sect.~\ref{methodsect} below).  We therefore use stringent cuts on both \texttt{sharp}, which robustly eliminates artifacts due to diffraction spikes from bright sources, as well as \texttt{crowd}.  Our \texttt{sharp} and \texttt{crowd} cuts are illustrated in Fig.~\ref{qparam_fig}, and were selected by comparing their distributions (versus magnitude) from our photometric catalog (after removing sources failing the other quality cuts) to their distributions from the output of artificial star tests (see below), which use only sources with stellar PSFs as input.
For each parameter (i.e., \texttt{crowd} and \texttt{sharp} in each of the two filters), an inflection point can be seen in the observed distributions (left column of each panel in Fig.~\ref{qparam_fig}) where non-stellar detections occupy loci that are unoccupied by recovered artificial stars.  We also use a looser cut on \texttt{crowd} in the chosen LW filter (F322W2) as compared to the SW filter (F150W) \citep[e.g.,][]{mcquinn24}, as the LW PSF is nearly double the width of the SW PSF for our chosen filters\footnote{\url{https://jwst-docs.stsci.edu/jwst-near-infrared-camera/nircam-performance/nircam-point-spread-functions}} .  

\begin{figure}
\gridline{\fig{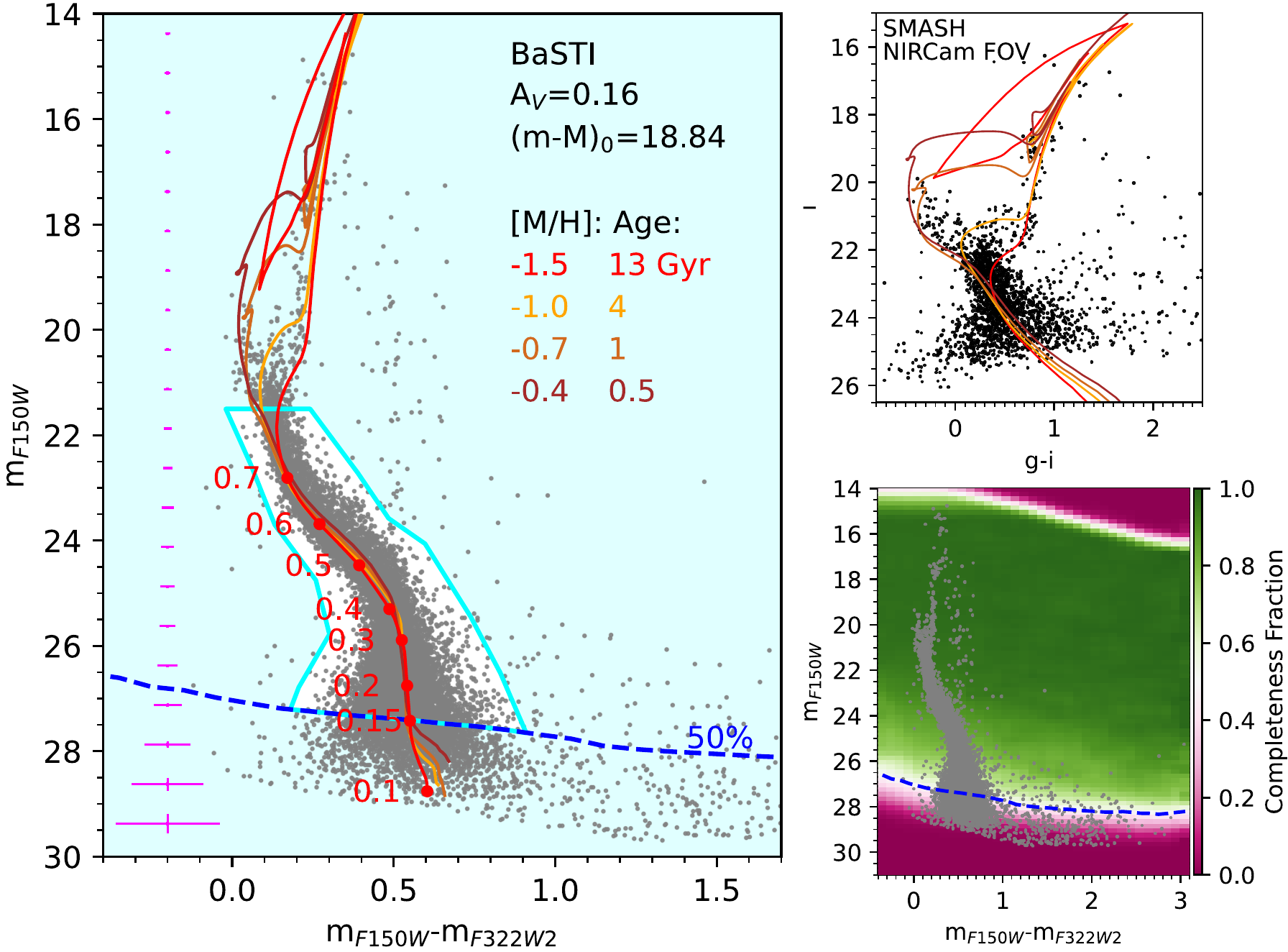}{0.99\textwidth}{}}
\caption{\textbf{Left:} CMD of all sources passing our photometric quality cuts.  Sources in the light blue shaded region were excluded from our IMF analysis to mitigate contamination by both Galactic halo sources (with m$_{\rm 150W}$-m$_{\rm F322W2}$$\sim$0.6 and 17$\lesssim$m$_{\rm 150W}$$\lesssim$24) and background galaxies appearing as faint red point sources (see Sect.~\ref{methodsect}).  Photometric errors in magnitude bins are shown along the left side in magenta, and the 50\% faint completeness limit (assessed from artificial star tests) is indicated by a dashed blue line.  BaSTI isochrones are overplotted assuming the extinction from \citet{skowron21} and mean line-of-sight distance from \citet{elyoussoufi21} to illustrate the lack of stars younger than $\sim$500 Myr.  Stellar masses (in M$_{\odot}$) are indicated along the main sequence of the ancient metal-poor isochrone, demonstrating that we achieve 50\% completeness at $\sim$0.16 M$_{\odot}$.  \textbf{Top Right:} CMD of sources from SMASH in our NIRCam field of view.  Isochrones are shown with the same representative ages and metallicities as in the left panel, confirming the lack of young ($\lesssim$500 Myr) stars.  \textbf{Bottom Right:} A map of photometric completeness over the CMD assessed from the artificial star tests.  The 50\% completeness limit and our observed catalog are shown as in the left panel.     
\label{cmd_fig}}
\end{figure}

We perform artificial star tests to quantify incompleteness and provide a noise model for application to synthetic photometry in our analysis.  Over 10$^{6}$ artificial stars were inserted in our images with a flat spatial distribution.  The color-magnitude distribution for 80\% of the artificial stars was based on the observed color distribution of sources passing our quality cuts, with the remaining 20\% of input artificial stars having a flat color-magnitude distribution to ensure coverage of the entire CMD (e.g., \citealt{gennaro18b}).  Artificial stars were subjected to the same PSF photometry procedure as real sources and considered recovered if they passed all of the photometric quality cuts described above.  A map of photometric completeness over the CMD, assessed from the artificial stars, is shown in the lower right-hand panel of Fig.~\ref{cmd_fig}.

\section{Analysis \label{anasect}}

To constrain the IMF of our target field, we use a forward modeling approach, generating synthetic photometry by sampling over the IMF parameters and comparing the resulting synthetic CMD with our observational catalog.  However, before performing this comparison, two additional steps are necessary, detailed below.  First, we impose color-magnitude cuts to restrict our IMF analysis to the main sequence of the SMC target population and minimize the impact of contaminants (Sect.~\ref{contamsect}).  Second, we correct our sample within this CMD region for luminosity-dependent color differences between observed and model-predicted colors (Sect.~\ref{isocorr_sect}).  Our synthetic photometry is then described in Sect.~\ref{inputsect}, followed by our fitting technique in Sect.~\ref{methodsect}.

\subsection{Minimizing Foreground and Background Contamination \label{contamsect}}

We restrict our IMF fitting to the CMD region shown in white in the left-hand panel of Fig.~\ref{cmd_fig}.  Operationally, any such CMD region can be selected for analysis precisely and arbitrarily by considering as undetected all artificial stars with output magnitudes outside the chosen region (importantly, no restriction is placed on the \textit{input} magnitudes of the artificial stars).  In the present case, we select a CMD region bounded at the faint end by the 50\% photometric completeness limit assessed from the artificial stars (although our results were ultimately insensitive to this choice; see Appendix \ref{testsect}).  At the bright end, our CMD region excludes post-main-sequence evolutionary products.  This choice minimizes the impact of any age dependency in the mass-luminosity relation, while also avoiding CMD regions where corrections between observed and model-predicted colors could depend on stellar evolutionary status for sources that are co-located on the CMD (e.g., for red giant, red clump and asymptotic giant branch stars at m$_{F150W}$$\lesssim$18.5), which is unknown a priori.  More generally, our chosen CMD region is selected to minimize contamination by both foreground Galactic stellar populations and background galaxies \citep[e.g.,][]{geha13,gennaro18a,gennaro18b,filion24}, which we now address in turn.  

\subsubsection{Foreground Contaminants \label{fgsect}}

We quantify contamination by foreground Galactic stellar populations using the TRILEGAL Galaxy model \citep{trilegal}.  Querying an area of 1 deg$^{2}$ centered on our NIRCam pointing, we perform 1000 monte carlo simulations in which the number of TRILEGAL-provided sources corresponding to our NIRCam field of view is drawn in each iteration, and the artificial star tests are used to apply photometric errors and incompleteness.  Over these iterations, TRILEGAL predicts 18$^{+4}_{-3}$ contaminants over the entire CMD region we use for our IMF analysis, or a contamination fraction of 0.1\%.  In Fig.~\ref{contam_fig} we illustrate how the fraction of foreground contaminants varies over the CMD, showing that they primarily populate a vertical sequence at (m$_{\rm F150W}$-m$_{\rm F322W2}$)$\sim$0.6 with 18$\lesssim$ m$_{\rm F150W}$$\lesssim$24 (comprised mostly of ancient halo dwarfs).  As a check on the TRILEGAL prediction, we select a CMD region that we expect to be populated only by Galactic foreground sources (i.e., devoid of SMC sources), shown as a dashed blue box in Fig.~\ref{contam_fig}.  We detect 41 sources in this box compared to 28$^{+5}_{-4}$ predicted by the Galaxy model (after accounting for observational noise), a generic consequence of the highly structured nature of the halo \citep[e.g.,][]{bell08}.  Even after accounting for this underprediction of foreground sources by the Galaxy model, the foreground contamination fraction in the CMD region we use for our IMF analysis (shown in grey in Fig.~\ref{contam_fig}) remains $<$0.2\%.

\subsubsection{Background Contaminants \label{bgsect}}

Despite the use of stringent photometric quality cuts, our photometric catalog may be contaminated by background galaxies that are sufficiently compact and/or distant to appear as point sources.  Ideally, such background galaxy contamination can be assessed using deep imaging of relatively \enquote{blank} high-latitude fields obtained with a similar observing setup \citep[e.g.,][]{mihos18,cohen18,cohen20}.  While such co-spatial imaging of sufficient depth in both of our chosen NIRCam filters (F150W and F322W2) is currently unavailable, we exploit the multi-wavelength catalogs provided by the CAnadian NIRISS Unbiased Cluster Survey (CANUCS; \citealt{sarrouh25}) program.  Although several other programs provide deep, multi-band NIRCam photometry targeting distant galaxies \citep[e.g.,][and references therein]{merlin24}, for our purposes CANUCS has two advantages: First, they provide multi-band catalogs from NIRCam imaging of five separate \enquote{NIRCam flanking fields} (NFFs, observed in parallel with their primary lensing fields), improving number density statistics.\footnote{Unlike the CANUCS primary fields, lensing magnification towards the NFFs is negligible \citep{asada25}.}  Second, they provide fluxes in as many as 11 NIRCam medium and wide filters in the LW channel alone, optimizing F322W2 fluxes calculated by interpolating in the available bandpasses (the F150W filter was included in CANUCS NFF imaging for all five pointings).  Further details on our usage of the CANUCS NFF catalogs are provided in Appendix \ref{bgcatsect}.  

\begin{figure}
\gridline{\fig{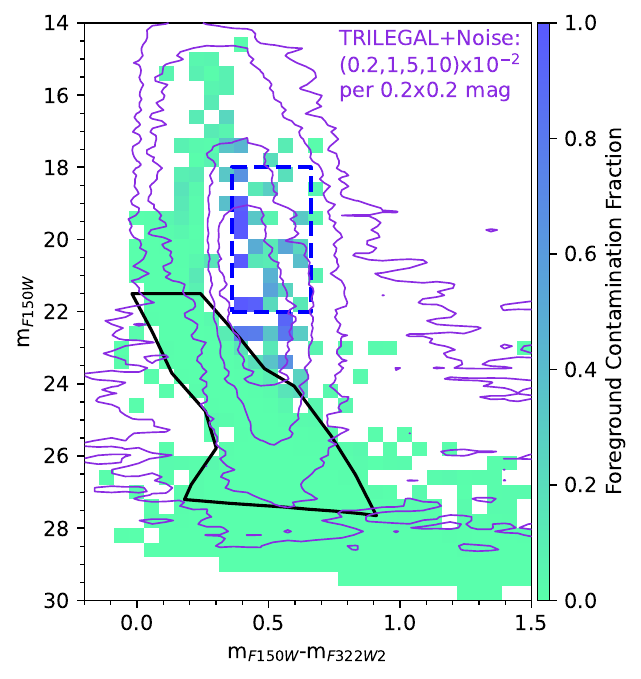}{0.43\textwidth}{}
          \fig{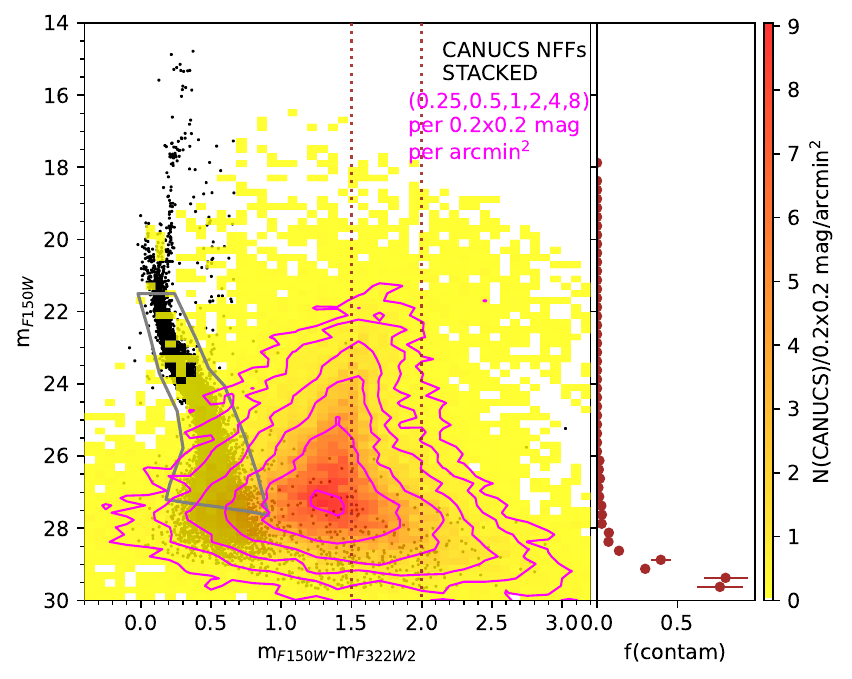}{0.56\textwidth}{}}
\caption{\textbf{Left:} Fractional contamination by Galactic foreground sources as a function of CMD location.  Purple contours show the density of foreground contaminants as indicated in the upper right corner.  The black area indicates the CMD region we use for our IMF analysis, avoiding CMD loci with statistically significant foreground contamination fractions.  The CMD region indicated by the dotted blue rectangle, devoid of true SMC sources, is used as a check on the model-predicted foreground contamination (see Sect.~\ref{fgsect}).  \textbf{Right:} The CMD distribution of background galaxies from the CANUCS NIRCam flanking fields (NFFs).  Again, the contamination fraction, plotted versus m$_{\rm F150W}$ in the rightmost panel, is estimated by comparing the number of observed versus predicted sources in a CMD region devoid of true SMC stellar detections (1.5$\leq$m$_{\rm F150W}$$\leq$2.0), indicted by dotted brown vertical lines.
\label{contam_fig}}
\end{figure}

The stacked CMD of the five CANUCS NFFs is shown in the right-hand panel of Fig.~\ref{contam_fig} (after applying our target field foreground extinction of A$_{V}$=0.16; see Sect.~\ref{methodsect}).  The majority of background sources are concentrated well redward of the SMC stellar evolutionary sequences, near (m$_{\rm F150W}$$-$m$_{\rm F322W2}$)$\sim$1.4.  To estimate the fraction of background galaxy contaminants, we again chose a color range where no true SMC sources are expected (1.5$\leq$m$_{\rm F150W}$-m$_{\rm F322W2}$$\leq$2.0), indicated by the dotted vertical lines in the right-hand panel of Fig.~\ref{contam_fig}.  We then compared the number of detected sources (i.e., background galaxies that survived our photometric quality cuts) to the total number observed in the CANUCS NFFs (normalized to our target field area).  The estimated contamination fraction of CANUCS sources is shown in the rightmost panel of Fig.~\ref{contam_fig}.  
The contamination fraction f(contam) rises sharply moving faintward towards our detection limit, but in the magnitude range used for our IMF analysis, background galaxy contamination is 1\% overall.
This procedure implicitly assumes that the depth and noise properties of  the galaxy fields are similar to our target field, although this approximation is quite reasonable: For the five CANUCS NFFs, \citet{sarrouh25} quote 3-$\sigma$ depths of 28.90$\leq$m$_{\rm F150W}$$\leq$29.15 
compared to our S/N=4 at m$_{\rm F150W}$$\sim$28.5.  At longer wavelengths the CANUCS 
3-$\sigma$ depths range from 27.64$\leq$m$_{\rm F322W2}$$\leq$28.20, 
surpassing our S/N=4 cutoff for (m$_{\rm F150W}$-m$_{\rm F322W2}$)$\gtrsim$1.4 where typical background galaxies lie.

\subsection{Correcting for Model-Data Color Differences \label{isocorr_sect}}

\begin{figure}
\gridline{\fig{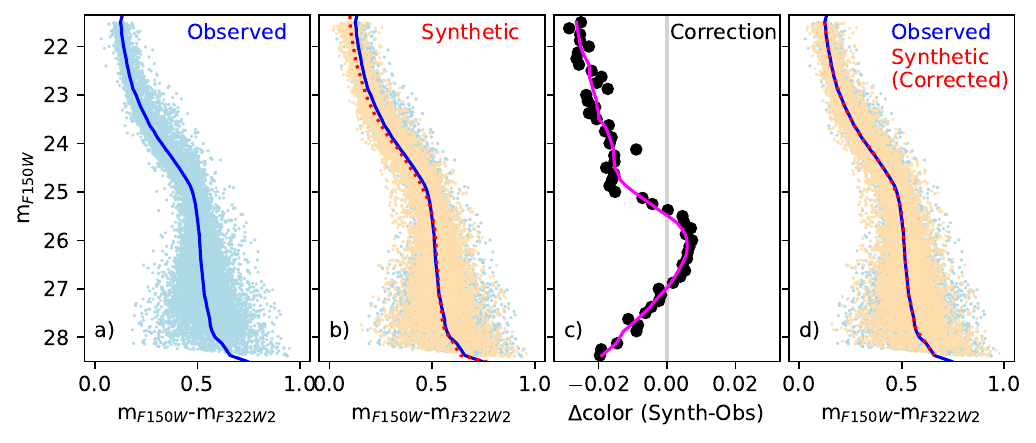}{0.99\textwidth}{}}
\caption{Illustration of our correction for the color difference between observed and model-predicted color.  The observed fiducial sequence (blue; panel a) is compared to the fiducial sequence obtained from synthetic photometry (red; panel b), and the difference (magenta; panel c) is applied to the synthetic photometry, with the results shown in panel (d).
\label{isocorr_fig}}
\end{figure}

In Fig.~\ref{isocorr_fig}, we compare our photometric catalog to model-predicted synthetic photometry, generated from the Bag of Stellar Tracks and Isochrones (BaSTI) stellar evolutionary models \citep{hidalgo18}.  In generating synthetic photometry, we use the input assumptions below in Sect.~\ref{inputsect}, applying observational noise and incompleteness assessed from the artificial stars.  Magnitude-dependent color offsets between the observed and synthetic photometry are small but non-negligible, possibly due at least in part to remaining uncertainties in stellar evolutionary models and color-temperature transformations.  We quantify and correct for these offsets by generating fiducial sequences for both the observed and synthetic CMDs, calculating the median 3-sigma-clipped color in bins of m$_{F150W}$ (shown in Figs.~\ref{isocorr_fig}a and \ref{isocorr_fig}b respectively).  The color differences between the observed and synthetic fiducial sequences are shown in Fig.~\ref{isocorr_fig}c, and the final corrections were obtained by smoothing with a Savitsky-Golay polynomial filter (shown in magenta).  The variation in color difference between the observed and predicted sequences as a function of magnitude is not well represented by a simple shift or low-order function.  Such a difference in the \textit{shape} of the fiducial sequences argues against a mismatch in distance and/or extinction as the cause of the offset.  Similarly, because our analysis is performed on the main sequence where color and magnitude vary relatively smoothly as a function of age and metallicity, the observed shape difference cannot be easily reconciled by changes in the assumed star formation rate or chemical evolution history.  For our downstream analysis, we apply the color shifts as a function of apparent magnitude when fitting for our IMF parameters by comparing observed and synthetic photometry.  

\subsection{Synthetic Photometry \label{inputsect}}

To constrain the IMF parameters towards our target field, we compare our photometry (over the selected CMD region; see Sect.~\ref{contamsect}) to synthetic photometry generated from BaSTI isochrones.  This synthetic photometry is generated exploiting numerous independent literature constraints on the properties of the stellar population towards our target field: 

\begin{itemize}

\item \textbf{Stellar line-of-sight distance distribution:} Any analyses of stellar populations towards the SMC are complicated by a line-of-sight depth that is significant and spatially variable \citep[e.g.,][]{nidever13,sub17,omkumar21,elyoussoufi21,tatton21}.  We assume the line-of-sight depth distribution from \citet[][see their Table 3]{elyoussoufi21}, based on double Gaussian fits to the red clump magnitude distribution in four azimuthal sectors of one-degree wide spatial annuli centered on the SMC.  For the location of our target field, their best-fit double Gaussian has a distance modulus to the more nearby Gaussian peak of $\mu_{1}$=18.79 mag, a difference between the two means of $\mu_{2}$-$\mu_{1}$ = 0.22 mag (corresponding to a distance difference of $\Delta$D=6.1 kpc), 
standard deviations of the two Gaussian peaks ($\sigma_{1}$, $\sigma_{2}$) = (0.29, 0.11) mag and relative amplitudes of (A$_{1}$/A$_{2}$) = 0.82, with relative uncertainties $\lesssim$5\% on all of these quantities.  These parameters produce a distance distribution in reasonable agreement with other recent literature estimates \citep{omkumar21,tatton21}.

\item \textbf{Extinction:} Estimates of the extinction towards our target field are provided by the map of \citet{skowron21}, built from optical photometry of the red clump throughout the LMC and SMC.  
They calculated the mean extinction and also fit asymmetric Gaussians allowing for different standard deviations on the near and far side of the SMC.    
Adopting their values for our target field location and coefficients from \citet{sf11}\footnote{While variability has been found in the extinction law towards the SMC, this has a negligible impact on our results since $E(F322W2-F150W)$/A$_{V}$$\approx$0.136 \citep[e.g.,][]{parsec}.  In any case, recent results find that the \textit{average} value of R$_{V}$ is similar to that of the Milky Way based on 32 sightlines towards the inner SMC \citep{gordon24}.} gives a mean extinction of A$_{V}$=0.16 mag (identical to the recent map of \citealt{oden25}) and a standard deviation of $\sigma$(A$_{V}$)=0.16 mag (corresponding to A$_{\rm F150W}$=0.037 mag,A$_{\rm F322W2}$=0.010 mag) along the line-of-sight.  We adopt a lognormal line-of-sight distribution for the extinction \citep[cf.][see their Sect.~3.1.4]{gennaro20}.

\item \textbf{Star Formation History:} We assume the star formation history (SFH) fit to the cospatial smc0100 field by \citet{noel09} using then-recent BaSTI stellar evolutionary models.  In practice, an assumed SFH is supplied as a probability distribution function and therefore requires no assumptions on the normalization of the star formation rate, which can depend on the IMF.  The cumulative mass fraction produced as a function of lookback time based on our assumed SFH is shown in the top panel of Fig.~\ref{sfhfig}.

\item \textbf{Chemical Evolution History:} As with the SFH, the evolution of metallicity with time can be supplied as a probability distribution.  We assume an age-metallicity relation (AMR) consisting of a linear relation between [M/H] and lookback time with an assumed spread of $\sigma$[M/H]=0.2 dex at fixed age, based on the AMR obtained for the colocated smc0100 field by \citet{carrera08}, shown in the lower panel of Fig.~\ref{sfhfig} (comparisons to spectroscopic metallicities of RGB stars are discussed in Appendix \ref{testsect}).  

\begin{figure}
  \begin{minipage}[c]{0.5\textwidth}
    \includegraphics[width=\textwidth]{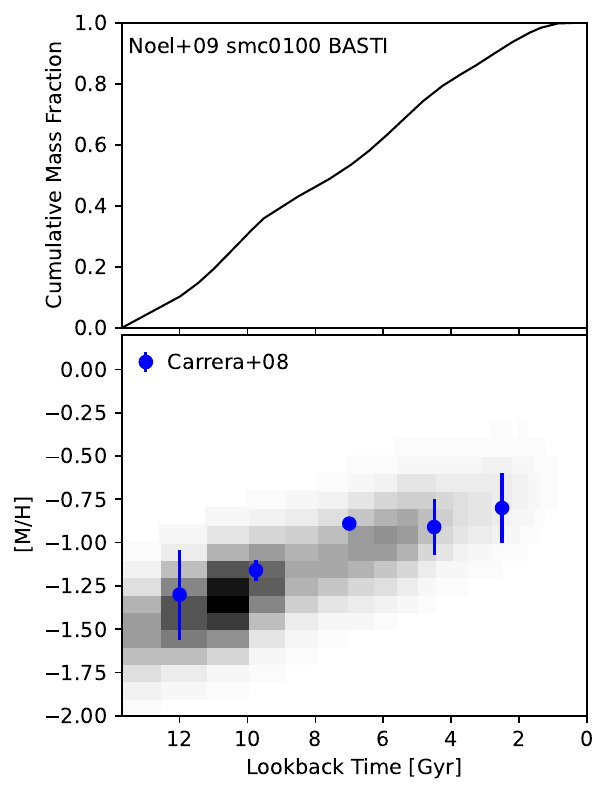}
  \end{minipage}\hfill
  \begin{minipage}[c]{0.45\textwidth}
    \caption{\textbf{Top:} The cumulative SFH of the co-spatial smc0100 field (see Fig.~\ref{fovfig}) presented by \citet{noel09}, which we assume when generating synthetic photometry for our IMF fitting.  \textbf{Bottom:} The age-metallicity relation we assume, provided as a probability distribution.  Darker shading indicates higher probabilities, noting that the total contribution in each interval of lookback time (horizontal axis) is set by the mass fraction formed in that time interval (upper panel).  Values from sources in the co-spatial smc0100 field \citep{carrera08} are overplotted. \label{sfhfig}}
\end{minipage}
\end{figure}

\end{itemize}

\subsection{Fitting Methodology \label{methodsect}}

Our technique for constraining the IMF parameters towards our target field is based on the Approximate Bayesian Computation Markov Chain Monte Carlo (ABC-MCMC) approach detailed in \citet{gennaro18a}, and implemented (including improvements described below) using the STARWAVE code.\footnote{\url{https://www.github.com/Resolved-Stellar-Populations/starwave}}  
Briefly, for a given set of trial IMF parameters, a synthetic CMD is generated under the assumptions listed in Sect.~\ref{inputsect}, folding in the noise properties of our observations (photometric error, bias and incompleteness) using our ensemble of $>$10$^{6}$ artificial stars.  For each set of trial parameters, a synthetic CMD is compared with the observed CMD using a kernel distance metric.  Here, we update the ABC-MCMC approach using simulation based inference\footnote{\url{https://github.com/sbi-dev/sbi}}, retaining the kernel distance metric as the goodness-of-fit indicator.  By using sequential neural posterior estimation (SNPE-C; \citealt{greenberg19}), a neural network is trained over multiple rounds to produce improved estimates for the posterior distributions of each parameter using an iterative procedure.  Initially, the neural network is trained using trial sets of parameter values drawn from the prior distributions.  Next, in subsequent rounds, the network is re-trained using trial parameter distributions drawn from the posteriors in the previous round, and we perform this iterative re-training over a total of five rounds with 10,000 simulations per round.  Validation of this approach using simulations is presented in Appendix \ref{validsect}.  

Along with posterior distributions for the IMF parameters, we also obtain a posterior distribution of the total number of stars N(tot), sampled over a broad, flat prior in Log N(tot) over the entire modeled mass interval (in this case 0.1$\leq$M/M$_{\odot}$$\leq$8, limited at the low-mass end by the masses provided with the evolutionary models; see \citealt{gennaro18a} for a discussion), which essentially functions as a normalization factor constrained by the total number of observed sources.  We also float the binary fraction f(bin) in our fitting, sampling over a flat prior 0$<$f(bin)$<$1, where each source is assumed to be either single or binary with a flat mass ratio distribution assumed for binaries.  

A novel aspect of our approach is that we also float distance and extinction properties towards our field as free parameters in our fit, allowing the posterior distributions of the IMF parameters to reflect their uncertainties as constrained by our CMD.  This is in contrast to several other recent IMF studies, which either assume fixed values of distance and extinction \citep[e.g.,][]{gennaro18a,gennaro18b} or fixed Gaussian distributions of their uncertainties \citep[e.g.,][]{filion22,filion24}.  Rather, we float the distance zeropoint of our assumed line-of-sight distance distribution ($\mu_{1}$, the distance to the nearer of the two Gaussian peaks along the line of sight) as well as the mean extinction and its lognormal width (A$_{V}$ and $\delta$A$_{V}$).  Regarding the distance distribution, we float only its zeropoint (parameterized by $\mu_{1}$) and hold the other parameters governing the \textit{shape} of the distribution fixed because we found that our data were insufficient to simultaneously disentangle these individual line of sight distance parameters (i.e., the widths and relative amplitudes of the two Gaussians and their separation) from the other free parameters (the IMF parameters, f(bin) and N(tot), even when fixing A$_{V}$ and $\delta$A$_{V}$ to their literature values given in Sect.~\ref{inputsect}).  However, we demonstrate in Appendix \ref{testsect} that adoption of an alternate literature distance distribution does not affect the IMF parameters beyond their 1-$\sigma$ uncertainties.  

We fit for three different types of mass distributions to describe the underlying IMF, a single power law (SPL), broken power law (BPL) and lognormal that transitions to a \citet{salpeter} power law for masses $m$$\geq$1M$_{\odot}$ (LN): 
\begin{equation}
p_{\rm SPL} (m \mid \alpha)\propto m^{\alpha}
\end{equation}

\begin{equation}
p_{\rm BPL} (m \mid \alpha_{1},\alpha_{2})\propto m^{\alpha_{1}}, m < 0.5M_{\odot}
\end{equation}
\begin{equation}
p_{\rm BPL} (m \mid \alpha_{1},\alpha_{2})\propto m^{\alpha_{2}}, m \geq 0.5M_{\odot}
\end{equation}

\begin{equation}
p_{\rm LN} (m \mid m_{c},\sigma)\propto \frac{1}{m}e^{-\frac{1}{2}(\frac{log(m)-log(m_{c})}{\sigma})^{2}}, m < 1M_{\odot}
\end{equation}
\begin{equation}
p_{\rm LN} (m)\propto m^{-2.35}, m \geq 1M_{\odot}
\end{equation}

Accordingly, in addition to f(bin), Log N(tot), $\mu_{1}$, A$_{V}$ and $\delta$A$_{V}$, the SPL form of the IMF has one additional free parameter (the IMF slope $\alpha$).  The BPL and LN forms of the IMF have two additional free IMF parameters each.  For the BPL fit, $\alpha_{1}$ and $\alpha_{2}$ represent the low-mass and high-mass power law IMF slopes, with the break between them fixed at $m_{b}$=0.5M$_{\odot}$ \citep{kroupa01,kroupa13}, although allowing $m_{b}$ to float as a free parameter in our fits did not significantly impact our results (see Sect.~\ref{bplsect} below).  For the lognormal IMF, $m_{c}$ and $\sigma$ represent the characteristic mass and lognormal width of the IMF.  

\section{Results \label{resultsect}}

\begin{figure}
\gridline{\fig{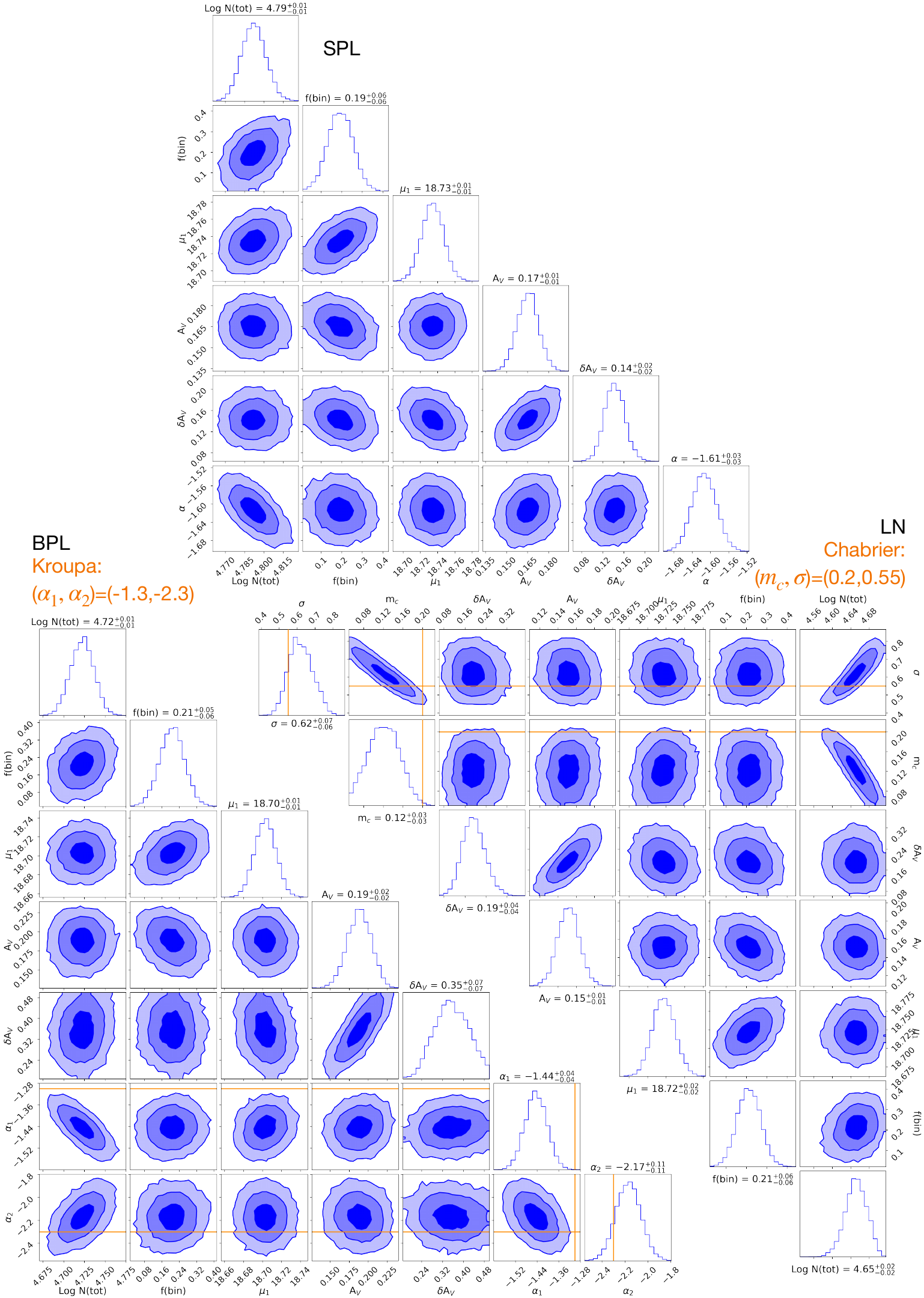}{0.835\textwidth}{}}
\vspace{-0.9cm}
\caption{Corner plots showing the posterior distributions of free parameters for each of the three functional forms of the IMF explored: single power law (SPL; top), broken power law (BPL; lower left) and lognormal (LN; lower right).  For the BPL and LN fits, canonical values from the solar neighborhood \citep{kroupa01,chabrier05} 
are indicated in orange and overplotted (the best-fit SPL slope differs from the \citet{salpeter} value by well over 3-$\sigma$, see Table \ref{imfpartab}).
\label{corner_fig}}
\end{figure}

The posterior distributions of best-fitting IMF parameters for each of the three analytical IMF forms explored (SPL, BPL, LN) are shown in Fig.~\ref{corner_fig}.  Contours represent 68\%, 95\% and 99.7\% confidence intervals\footnote{For a two-dimensional distribution, these correspond to 39.4\%, 86.5\% and 98.9\% respectively} (which we also refer to as 1, 2 and 3-$\sigma$), tabulated along with the median values in Table \ref{imfpartab}.  In Fig.~\ref{corner_fig}, solar neighborhood values for SPL, BPL and LN fits, from \citet{salpeter}, \citet{kroupa01} and \citet{chabrier05} respectively, are also overplotted in orange for comparison.   

\begin{deluxetable}{lcccc}
\tablecaption{IMF Parameters\label{imfpartab}}
\tablehead{
\colhead{Parameter} & \colhead{Median} & \colhead{$\pm$1-$\sigma$} & \colhead{$\pm$2-$\sigma$} & \colhead{$\pm$3-$\sigma$}}
\startdata
\hline
\multicolumn{5}{c}{Single power law:} \\
\hline
Log N(tot) & 4.79 & 4.78 4.80 & 4.77 4.81 & 4.77 4.82 \\
f(bin) & 0.19 & 0.13 0.26 & 0.07 0.32 & 0.02 0.37 \\
$\mu_{1}$ & 18.73 & 18.72 18.75 & 18.71 18.76 & 18.70 18.77 \\
A$_{V}$ & 0.17 & 0.16 0.17 & 0.15 0.18 & 0.14 0.19 \\
$\delta$A$_{V}$ & 0.14 & 0.12 0.16 & 0.11 0.18 & 0.09 0.20 \\
$\alpha$ & -1.61 & -1.64 -1.59 & -1.67 -1.56 & -1.69 -1.54 \\
\hline
\multicolumn{5}{c}{Broken power law:} \\
\hline
Log N(tot) & 4.72 & 4.71 4.74 & 4.69 4.75 & 4.68 4.76 \\
f(bin) & 0.21 & 0.15 0.26 & 0.09 0.32 & 0.04 0.37 \\
$\mu_{1}$ & 18.70 & 18.69 18.71 & 18.68 18.73 & 18.67 18.74 \\
A$_{V}$ & 0.19 & 0.17 0.20 & 0.16 0.22 & 0.14 0.23 \\
$\delta$A$_{V}$ & 0.35 & 0.29 0.42 & 0.23 0.48 & 0.18 0.50 \\
$\alpha_{1}$ & -1.44 & -1.49 -1.40 & -1.53 -1.36 & -1.56 -1.32 \\
$\alpha_{2}$ & -2.17 & -2.28 -2.06 & -2.38 -1.95 & -2.50 -1.85 \\
\hline
\multicolumn{5}{c}{Lognormal:} \\
\hline
Log N(tot) & 4.65 & 4.63 4.68 & 4.60 4.69 & 4.57 4.71 \\
f(bin) & 0.21 & 0.16 0.27 & 0.10 0.33 & 0.05 0.38 \\
$\mu_{1}$ & 18.72 & 18.71 18.74 & 18.69 18.75 & 18.68 18.77 \\
A$_{V}$ & 0.15 & 0.14 0.16 & 0.13 0.18 & 0.11 0.19 \\
$\delta$A$_{V}$ & 0.19 & 0.15 0.24 & 0.11 0.29 & 0.08 0.34 \\
m$_{c}$ & 0.12 & 0.09 0.15 & 0.06 0.18 & 0.05 0.21 \\
$\sigma$ & 0.62 & 0.55 0.69 & 0.50 0.76 & 0.45 0.81 \\
\enddata
\tablecomments{1, 2 and 3-$\sigma$ intervals correspond to 68\%, 95\% and 99.7\% confidence intervals assessed from the marginal posterior distributions of each parameter.}
\end{deluxetable}

There are two aspects of the posterior distributions in Fig.~\ref{corner_fig} that hold true regardless of the assumed functional form of the IMF: First, the IMF parameters can be degenerate with the total number of stars N(tot).  This is a natural consequence of the need to produce a fixed number of observed sources in the chosen CMD region while changing the underlying stellar mass distribution \citep[e.g.,][]{gennaro18a}.  
Second, we find binary fractions of f(bin)=0.2 to within 1-$\sigma$ confidence limits, and f(bin)$<$0.4 at more than 3-$\sigma$ confidence, for all three IMF parameterizations.  This is consistent with main sequence binary fractions that decrease with primary mass as observed in the solar neighborhood \citep[][and references therein]{offner23}, as well as f(bin) measured in the Coma Berenices UFD using a sample extending nearly as deep as ours (down to $\sim$0.23M$_{\odot}$; \citealt{gennaro18b}).  While observations within the Milky Way indicate that the decrease in f(bin) with primary mass may be counteracted to some extent by an \textit{increase} in close binary fraction at lower metallicities \citep[e.g.,][]{moe19,mazzola20}, additional constraints (i.e., on the mass ratio distribution) are necessary to evaluate whether this conclusion can be generalized to external galaxies (see Sect.~\ref{constraintsect}).  

\begin{figure}
\gridline{\fig{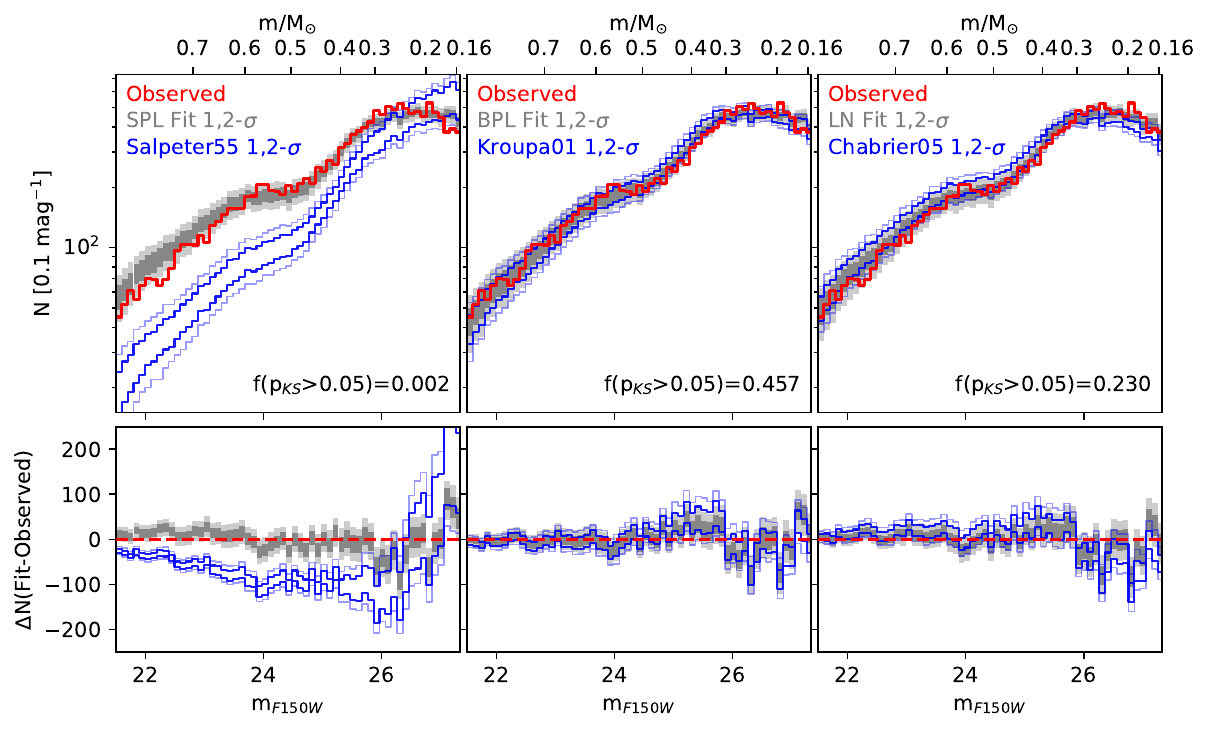}{0.99\textwidth}{}}
\vspace{-0.5cm}
\caption{\textbf{Top Row:} Our observed m$_{F150W}$ luminosity functions (LFs) are shown in red for each of the three IMF parametizations explored (from left to right: SPL, BPL and LN), binned in intervals of 0.1 mag for visualization purposes.  1 and 2-$\sigma$ uncertainty intervals are shown using dark and light grey shading respectively, evaluated from synthetic photometry using draws from the posterior parameter distributions.  In each panel, we give $f(p_{KS}>0.05)$, the fraction of posterior draws with $p$-values$>$0.05 from two-sided K-S tests comparing observed and synthetic LFs.  For comparison, the 1 and 2-$\sigma$ envelope of LFs constructed from synthetic photometry assuming solar neighborhood values of the IMF parameters \citep{salpeter,kroupa01,chabrier05} are shown in blue.  \textbf{Bottom Row:} Same, but showing the differences between the observed and synthetic LFs.  
\label{resids_fig}}
\end{figure}

In Fig.~\ref{resids_fig}, we display the observed m$_{\rm F150W}$ luminosity function (LF) for each of the three IMF parameterizations (SPL, BPL, LN) in red in the top row, binned for visualization purposes.  The 1 and 2-$\sigma$ uncertainty intervals are shown using dark and light grey shading respectively, calculated by constructing LFs from synthetic photometry over each of 1,000 draws from the posterior distributions of the IMF parameters.  For comparison, the 1 and 2-$\sigma$ intervals of LFs that would result from solar neighborhood values (shown in orange in Fig.~\ref{corner_fig}) are shown in blue.  The SPL fit is clearly flatter than the \citet{salpeter} value, as expected given the mass range we probe.  
The BPL and LN fits, shown in the middle and right-hand columns of Fig.~\ref{resids_fig}, show deviations from the observations that are only marginally significant ($\lesssim$2-$\sigma$), with the BPL fit appearing to provide the best match to the observed LF.  Fit quality across the three IMF parameterizations is quantified using two-sided Kolmogorov-Smirnov (K-S) tests, comparing synthetic m$_{F150W}$ LFs generated using 1,000 draws from the posterior distributions (shown in Fig.~\ref{corner_fig}) to the observed m$_{F150W}$ LF for each of the three IMF parameterizations (SPL, BPL, LN).  As in other recent implementations of this approach \citep{safarzadeh22,filion22,filion24}, we set a $p$-value threshold of $p$$<$0.05 to reject the null hypothesis (i.e., that the observed and synthetic LFs have the same parent distribution).  In each panel of Fig.~\ref{resids_fig} we report $f(p_{KS}>0.05)$, the fraction of posterior draws that resulted in $p$-values$>$0.05, confirming the improved quality of the BPL and LN fits compared to the SPL fits.  Specifically, the synthetic LFs and the observed LF were consistent with being drawn from the same parent distribution in only 0.2\% of cases for the SPL fits, compared to 45.7\% and 23.0\% for the BPL and LN fits respectively.  This result would be essentially unaffected had we required $p$-values$>$0.05 for \textit{both} the m$_{F150W}$ \textit{and} m$_{F322W2}$ LFs \citep[e.g.,][]{filion22,filion24}, yielding $f(p_{KS}>0.05)$=(0.001, 0.392, 0.203) for (SPL, BPL, LN) IMF parameterizations.  

We now turn to a quantitative comparison of our IMF parameters with other metal-poor galaxies as well as results obtained for the Milky Way (and its subpopulations).  

\section{Discussion: The SMC IMF in Context \label{discusssect}}
\subsection{Single Power Law Fits \label{splsect}}

\begin{figure}
\gridline{\fig{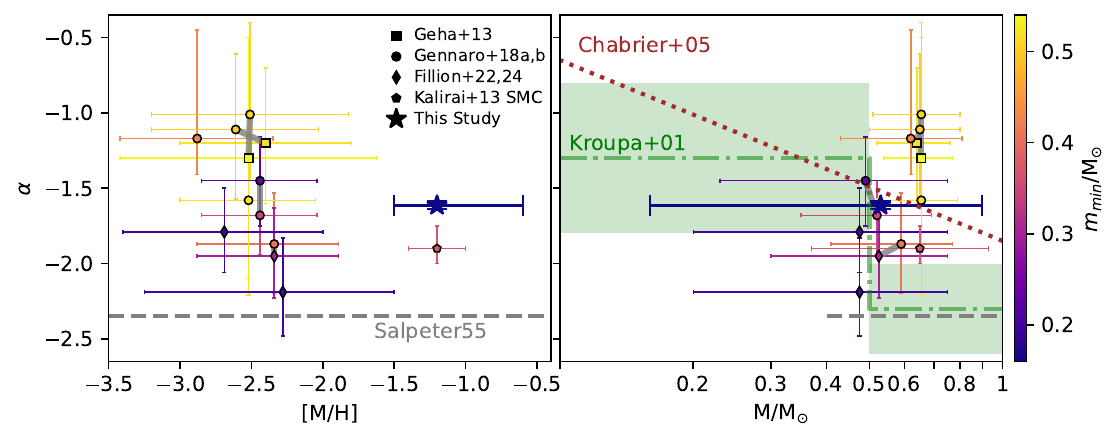}{0.99\textwidth}{}}
\caption{Comparison of single power law IMF slopes fit to our target field (shown as a star) versus other low-metallicity dwarf galaxies (with symbols indicating literature sources).  
Slopes are shown versus metallicity (left) and stellar mass range used to fit the IMF slope (right), color-coded by the mimimum mass (in M$_{\odot}$) of the stellar sample included in the IMF fits.  Results from mutiple studies of the same galaxy are connected with solid grey lines.  The \citet{salpeter} IMF slope is shown as a dashed grey line, and in the right-hand panel the broken power law slopes of \citet{kroupa01} and their uncertainties are shown in green, while the slope that would be obtained by fitting a single power law to a \citet{chabrier05} IMF (i.e., the slope of the tangent line to the lognormal IMF) as a function of stellar mass is shown as a brown dotted line.  The slope of our SPL fit is consistent with slopes found for all but one UFD analyzed using similarly deep imaging, while potential causes of the minor difference in slope ($\Delta$$\alpha$$\lesssim$0.3) versus the \citet{kalirai13} SMC field are discussed in the text.  
\label{spl_fig}}
\end{figure}

In the left panel of Fig.~\ref{spl_fig}, we compare our SPL slope to values measured in UFDs and the \citet{kalirai13} SMC field as a function of metallicity.   
Importantly, differences in the sampled mass range can affect the best-fit SPL slope by $\sim$0.2 or more in the case of an underlying IMF that is truly lognormal (e.g., Sect.~4.3 of \citealt{yan24}, also see reviews by \citealt{hennebelle24} and \citealt{kroupa24}), so points are color-coded by $m_{min}$, the minimum stellar mass probed by each IMF study.   
We find a slightly shallower slope than measured by \citet{kalirai13} from proper-motion-cleaned deep optical imaging of a field $\sim$2.6 kpc west of the SMC.  However, such a difference may be attributed at least in part to true differences in the underlying stellar populations properties between our field and their given an azimuthally-dependent radial metallicity gradient in the SMC \citep{navabi25}, as well as methodological differences including their assumptions of a higher binary fraction and a flat metallicity distribution and their use of fixed magnitude bins to fit an IMF to their IMF.  Meanwhile, compared to the UFDs with available IMF slopes, our single power law slope is consistent with all but one other UFDs measured to mass limits nearly as deep as ours ($m_{min}$$\lesssim$0.3M$_{\odot}$).  The exception is Ursa Major II, for which \citet{filion24} suggest background galaxy contamination could be a factor based on the steepening BPL slopes they recover at lower stellar masses (i.e. with a slope change in the opposite sense as a \citealt{kroupa01} IMF).  

Among the UFDs, there are three (Canes Venatici II, Hercules, Leo IV) that have SPL slopes flatter (i.e., less negative) than ours beyond 1-$\sigma$ confidence, and all of these UFDs also have slope measurements based on shallower, optical imaging ($m_{min}$$>$0.43M$_{\odot}$).  The tendency for targets to show shallower slopes at larger $m_{min}$ is opposite to what would be expected for an underlying lognormal mass function (e.g., the right-hand panel of Fig.~\ref{spl_fig}, also see \citealt{elbadry17}).  However, given the small sample size and use of shallow optical imaging, this trend could be an artifact of background contamination and/or variations in the binary fraction and/or mass ratio distribution, which remain poorly constrained in UFDs in most cases \citep[e.g.,][]{gennaro18a,gennaro18b,filion24}.  

Compared to SPL slopes measured in the solar neighborhood, the extent to which our results are consistent with previous studies varies, although as in the case of the UFDs and the \citet{kalirai13} SMC field, a direct comparison is hampered by differences in sample selection and methodology (we return to this point below in Sect.~\ref{constraintsect}).  
\citet{sollima19} report a single power law slope of $\alpha$=$-$1.34$\pm$0.07 over the mass range 0.11$\leq$M/M$_{\odot}$$\leq$0.93 assuming a single stellar population age of 10 Gyr and an (asymmetric) Gaussian metallicity distribution centered on solar.  In addition, several recent studies have used single power law fits to investigate variations in $\alpha$ within the Milky Way, separating various subpopulations chemodynamically.  For example, \citet{hallakoun21} present evidence that accreted metal-poor halo stars with -2$\geq$[M/H]$\geq$-0.6 are well represented by a single power law IMF, unlike the other Galactic stellar populations that they analyze (thin disk and thick disk).  They measure $\alpha$=$-$1.82$^{+0.14}_{-0.17}$ over a mass range of $\sim$0.2$\leq$M/M$_{\odot}$$\leq$0.9 for the metal-poor halo subsample, differing from our result by $<$2-$\sigma$ and in good agreement with other studies probing UFDs and the SMC to similarly low stellar masses (see Fig.~\ref{spl_fig}).  
In addition, single power law slopes over the range $-$2.8$\lesssim$$\alpha$$\lesssim$$-$2.2 were reported to vary as a function of metallicity ($-$0.8$\leq$[Fe/H]$\leq$+0.05) by \citet{li23}.  However, their IMF slopes were fit over a higher, more restricted mass range (0.3$\leq$M/M$_{\odot}$$\leq$0.7), at least partially explaining the discrepancy with the \citet{sollima19} value 
together with differences in sample selection and methodology, including the assumed mass-luminosity relation and binary fraction.  

More generally, high-quality IMF measurements, where available, indicate that a single power law can be a poor representation of the subsolar-mass IMF at various metallicities.  This has been evident in the solar neighborhood for decades \citep[e.g.,][also see \citealt{kirkpatrick24}]{scalo86,ktg93,kroupa01}, and is supported by two pieces of evidence in the SMC (i.e., at lower metallicity): First, our K-S tests indicate that SPL fits are a poor representation of the observed LFs compared to BPL or LN fits (see Sect.~\ref{resultsect}).  Second, a dropoff is observed at the faint end of the LF that is inconsistent with the best-fit synthetic SPL LFs (see the left panel of Fig.~\ref{resids_fig}).  Such a dropoff is also seen at the faint end of the LFs in the \citet{kalirai13} SMC field despite a detection limit $\sim$0.2M$_{\odot}$ higher (see their Fig.~5), suggesting that this may be a generic consequence of fitting a single power law slope to an underlying IMF that is better represented by a lognormal or broken power law.  

\subsection{Broken Power Law Fits \label{bplsect}}

The slopes of our broken power law fits ($\alpha_{1}$=$-$1.44$^{+0.04}_{-0.04}$,  $\alpha_{2}$=$-$2.17$^{+0.11}_{-0.11}$; see Table \ref{imfpartab}) are consistent with the empirically-motivated Galactic values of \citet{kroupa01} ($\alpha_{1}$=$-$1.3$\pm$0.5, $\alpha_{2}$=$-$2.3$\pm$0.3) to within their uncertainties.  This is also true of all UFDs where such fits were explored thus far \citep{gennaro18b,filion22,filion24}, with the possible exception of Reticulum II, for which \citet{safarzadeh22} find a LF inconsistent with  a \citet{kroupa01} IMF, although they assumed a single metallicity and binary properties extrapolated from projected separations in HST optical imaging.  

In the Milky Way, fits to chemodynamically selected subpopulations also yielded values of $\alpha_{1}$ and $\alpha_{2}$ consistent with \citet{kroupa01} values to within uncertainties in most cases \citep{hallakoun21}, with three marginal exceptions.  First, for their metal-poor halo subpopulation, \citet{hallakoun21} find ($\alpha_{1}$, $\alpha_{2}$) = ($-$1.51$\pm$0.06, $-$2.97$\pm$0.21) but over a slightly higher mass range (0.2$\lesssim$M/M$_{\odot}$$\lesssim$0.9), fixing the break mass between the two power law slopes at $m_{b}$=0.5M$_{\odot}$ in their fits as we have done.  Second, \citet{wang25} find Galactic values of ($\alpha_{1}$, $\alpha_{2}$) = ($-$0.81$^{+0.05}_{-0.06}$, $-$2.12$^{+0.04}_{-0.04}$), floating $m_{b}$ as a free parameter and recovering $m_{b}$=0.41$^{+0.01}_{-0.01}$M$_{\odot}$, producing an IMF consistent with \citet{kroupa01} to within uncertainties (see Fig.~8 of \citealt{wang25})\footnote{In the context, it also bears mention that the \citet{li23} sample, as a whole, is discrepant from the \citet{kroupa01} BPL IMF by $\lesssim$2$\sigma$ (see their Fig.~2).}.  While we have fixed $m_{b}$=0.5M$_{\odot}$ in our fits, we found that allowing $m_{b}$ to vary still produces IMF parameters consistent with \citet{kroupa01} values, albeit with less precision due to the covariance of $m_{b}$ with $\alpha_{1}$ and $\alpha_{2}$, finding ($\alpha_{1}$, $\alpha_{2}$, $m_{b}$/M$_{\odot}$) = ($-$1.31$^{+0.13}_{-0.08}$, $-$2.02$^{+0.21}_{-1.00}$, 0.49$^{+0.23}_{-0.13}$).  
Third, \citet{qiu25} report a trend between BPL power law slopes and metallicity in the solar neighborhood (150$\leq$D$_{\odot}$$\leq$350 pc) such that $\alpha_{1}$ and $\alpha_{2}$ become flatter for more metal-poor subsamples over the range $-$1$\leq$[Fe/H]$\leq$+0.5.  However, after accounting for unresolved binaries, a slightly different sampled mass range (0.25$\leq$M/M$_{\odot}$$\leq$1) as well as uncertainties related to their choice of $m_{b}$, assumed binary fraction, and use of binned mass intervals (each impacting their BPL slopes at a level of $\Delta$$\alpha$$\gtrsim$0.1; see their Figs.~10-13), our BPL slopes (see Fig.~\ref{corner_fig} 
and Table \ref{imfpartab}) are not inconsistent with their results for their most metal-poor subsample ($-$1.0$\leq$[Fe/H]$\leq$$-$0.8).

Lastly, we note that that the BPL fits yielded best-fit values of the internal extinction $\delta$A$_{V}$=0.35$^{+0.07}_{-0.07}$ mag (see Fig.~\ref{corner_fig} and Table \ref{imfpartab}), significantly higher than those found for the SPL or LN fits or predicted by the \citet{skowron21} reddening maps ($\delta$A$_{V}$$\approx$0.16 mag).  However, application of our fitting technique to synthetic data generated from a BPL IMF yielded a similarly offset posterior distribution for $\delta$A$_{V}$, but importantly, returned input IMF parameters ($\alpha_{1}$, $\alpha_{2}$) and binary fraction f(bin) to well within uncertainties (see Appendix \ref{validsect}).

\subsection{Lognormal Fits \label{ln_sect}}

Our lognormal fits (with best-fit parameters $m_{c}$=0.12$^{+0.03}_{-0.03}$M$_{\odot}$, $\sigma$=0.62$^{+0.07}_{-0.06}$M$_{\odot}$) are compatible with lower $m_{c}$ and higher $\sigma$ than the \citet{chabrier05} Galactic values (see Fig.~\ref{corner_fig}).  A similar trend was found for three of the five UFDs (Ursa Major II, Segue I, Bo{\"o}tes I) analyzed by \citet{filion22,filion24}, fitting resolved photometry only slightly shallower than ours (reaching $\leq$0.3M$_{\odot}$).  Interestingly, the trend of $m_{c}$ decreasing with metallicity is also seen in both observations of young Galactic stellar populations at subsolar metallicity \citep{yasui24,andersen25} as well as multiple simulations.  In the models of \citet{chon24}, stellar feedback impacts the IMF, heating dust grains and raising the temperature of gas near massive stars, suppressing fragmentation.  In addition, simulations by \citet{mathew25} find that the cloud virial parameter, which parameterizes the relationship between gravity and turbulence in star forming environments, may shift the IMF peak by as much as a factor of two, consistent with our results. 
On the other hand, three of the six UFDs analyzed by \citet{gennaro18a} have $m_{c}$ values that are higher (rather than lower) than Galactic values at more than 1$\sigma$ (but less than 2$\sigma$) confidence (and Milky Way-like values of $\sigma$$\gtrsim$0.55M$_{\odot}$).  Further clarification of the metallicity dependence of $m_{c}$ and $\sigma$ likely awaits ultra-deep imaging of UFDs, potentially within the reach of JWST (M. Gennaro et al., in prep., also see \citealt{shariat25}).

\subsection{Improving Constraints on the Low Mass IMF \label{constraintsect}}

Our results rely on several simplifying assumptions that were necessary based on available constraints, but may be improved with future observations.  For example, the properties of binaries towards the SMC remain poorly constrained at sub-solar masses.  
We have assumed a flat mass ratio distribution for binaries as in most previous IMF studies of metal-poor dwarfs \citep{geha13,kalirai13,gennaro18a,gennaro18b,filion22,filion24}, although high-quality observations in the Milky Way have revealed more complex trends \citep[e.g.,][]{moe17,moe19,elbadry19,offner23}.  

Comparisons to Milky Way IMF measurements are complicated by the fact that recent constraints on the local binary fraction and mass ratio distribution are handled differently (or not at all) across various recent measurements of the Galactic IMF \citep[e.g.,][]{sollima19,hallakoun21,li23,wang25}, exacerbated in some cases by the use of fixed mass bins when measuring the IMF and its uncertainties \citep{hallakoun21,yasui23,yasui24,andersen25,qiu25}, potentially affecting the results at a statistically significant level (\citealt{ma05,cara08}; see \citealt{ma09} for a review).  Looking forward, independent constraints on the binary properties of metal-poor populations may be provided by a combination of high spatial resolution imaging and/or temporally resolved radial velocities obtained from spectroscopy \citep{minor19,safarzadeh22,arroyo23,shariat25,qiu_bf}.  In the meantime, we note that our treatment of binaries as unresolved is reasonable given our target distance and available spatial resolution: Conservatively assuming the nearest of the best-fit $\mu_{1}$ values from Sect.~\ref{resultsect}, the minimum separable distance of two neighboring stars using our DOLPHOT PSF photometry parameters is $\sim$2600 AU.  This distance is beyond the separation of $>$90\% of Galactic F-M-type binaries and more than two orders of magnitude greater than the median separation of binaries with M$\leq$M$_{\odot}$ and [Fe/H]$\sim$$-$1 \citep{offner23}.

Our second simplifying assumption is that the mass function towards our target field has not been dynamically altered in any stellar-mass-dependent way.  Meanwhile, there is a vast body of chemodynamical evidence that the SMC is a complex, potentially multi-component system that is not well mixed due to recent interactions with the LMC \citep[e.g.,][]{besla12,zivick18,deleo20,mucciarelli23,murray24}.  Towards the eastern part of the SMC where our target field lies, chemodynamical studies targeting RGB stars reveal that the more nearby of the line-of-sight components is preferentially being pulled towards the LMC, and may have originated in the inner SMC \citep{almeida24}.    
However, the density of (CMD-selected) young stars towards our field was too low to assess any stellar-mass-dependent differences in kinematics \citep{niederhofer21}.  Fortunately, early studies quantifying the astrometric precision attained by JWST's imagers indicate that tests for mass-dependent kinematics well down the main sequence of the LMC and SMC may be well within reach shortly \citep[e.g.,][]{libralato23}.  Similarly, we have assumed that the stellar age and metallicity distributions towards our target field are independent of their line-of-sight distances.  While the symmetry between the kinematics of RGB and RC stars in both line-of-sight components argues against a gross asymmetry in stellar population properties along the line of sight \citep{james21}, deep multi-band imaging could provide per-star distances with the fidelity to test such an assumption (e.g., via SED fitting; \citealt{gordon16}; \citealt{lindberg25}).

\section{Conclusions \label{conclusion}}

We have targeted a single NIRCam field located $\sim$2.3 kpc south of the SMC main body (see Sect.~\ref{targetsect}, Fig.~\ref{fovfig}), exploiting independent measurements of the SFH, chemical evolution history, and extinction and line-of-sight distance distributions to measure the IMF down to $\sim$0.16M$_{\odot}$ (see Sect.~\ref{inputsect}).  After correcting for both detector-dependent photometric offsets in our imaging (Appendix \ref{zptsect}, Fig.~\ref{colorcorr_chip_fig}) as well as non-linear photometric offsets with respect to stellar evolutionary models (Sect.~\ref{isocorr_sect}, Fig.~\ref{isocorr_fig}), we used a CMD region minimizing foreground and background contaminants (Sects.~\ref{fgsect}-\ref{bgsect}, Fig.~\ref{contam_fig}) to constrain SPL, BPL and LN IMF parameters via forward modeling.  Our primary findings are: 

\begin{itemize}
\item The best-fit single power law slope (Sect.~\ref{splsect}) is $\alpha$=$-$1.61$^{+0.03}_{-0.03}$ (68\% confidence interval), statistically consistent with results for metal-poor UFDs ($\langle$[M/H]$\rangle$$<$$-$2) probed to similar depths ($m_{min}$$\lesssim$0.3M$_{\odot}$; see Table \ref{imfpartab} and Fig.~\ref{spl_fig}).  

\item We measure broken power law IMF slopes (Sect.~\ref{bplsect}) consistent with \citet{kroupa01} values, regardless of whether $m_{b}$ is held fixed.  When allowed to float as a free parameter, we recover $m_{b}$=0.49$^{+0.23}_{-0.13}$M$_{\odot}$, arguing against a value drastically different than the oft-assumed $m_{b}$=0.5 (see Fig.~\ref{corner_fig} and Sect.~\ref{bplsect}).  

\item Fits of a lognormal (LN) IMF (Sect.~\ref{ln_sect}) allow a characteristic mass $m_{c}$ (and lognormal width $\sigma$) lower (higher) than solar neighborhood values (Fig.~\ref{corner_fig}, Table \ref{imfpartab}), consistent with recent observations of young metal-poor clusters in the Milky Way and multiple simulations.  

\item Using two-sided K-S tests to assess fit quality, BPL and LN fits yield a better fit quality ($f(p_{KS}>0.05)$=45.7\% and 23.0\% respectively) than SPL fits, with $f(p_{KS}>0.05)$=0.2\% (Sect.~\ref{resultsect}, Fig.~\ref{resids_fig}).  

\item All three of the IMF parameterizations we examined (SPL, BPL, LN) yielded binary fractions f(bin)=0.2 to within 1-$\sigma$ confidence limits, consistent with the decrease in f(bin) with primary mass observed in the solar neighborhood.  

\end{itemize}

\begin{acknowledgements}

This work is based in part on observations made with the NASA/ESA/CSA James Webb Space Telescope.  The data were obtained from the Mikulski Archive for Space Telescopes at the Space Telescope Science Institute, which is operated by the Association of Universities for Research in Astronomy, Inc., under NASA contract NAS 5-01327 for JWST.  These observations are associated with program JWST-GO-2918.  The specific observations analyzed can be accessed via DOI 10.17909/sm2a-8w46.  
Support for this work was provided by NASA through grant GO-2918 from the Space Telescope Science Institute, which is operated by AURA, Inc., under NASA contract 5-26555.  This research has made use of the NASA Astrophysics Data System Bibliographic Services.  R.~E.~C. acknowledges support from Rutgers the State University of New Jersey.  

\end{acknowledgements}

\facilities{JWST (NIRCam)}
\software{astropy \citep{astropy}, matplotlib \citep{matplotlib}, numpy \citep{numpy}, sbi \citep{sbi}, DOLPHOT \citep{dolphot,dolphot2}.}

\bibliography{IMF_SMC}
\bibliographystyle{aasjournalv7}

\appendix

\section{Correcting Detector-Dependent Photometric Offsets \label{zptsect}}

\begin{figure}[h!]
\gridline{\fig{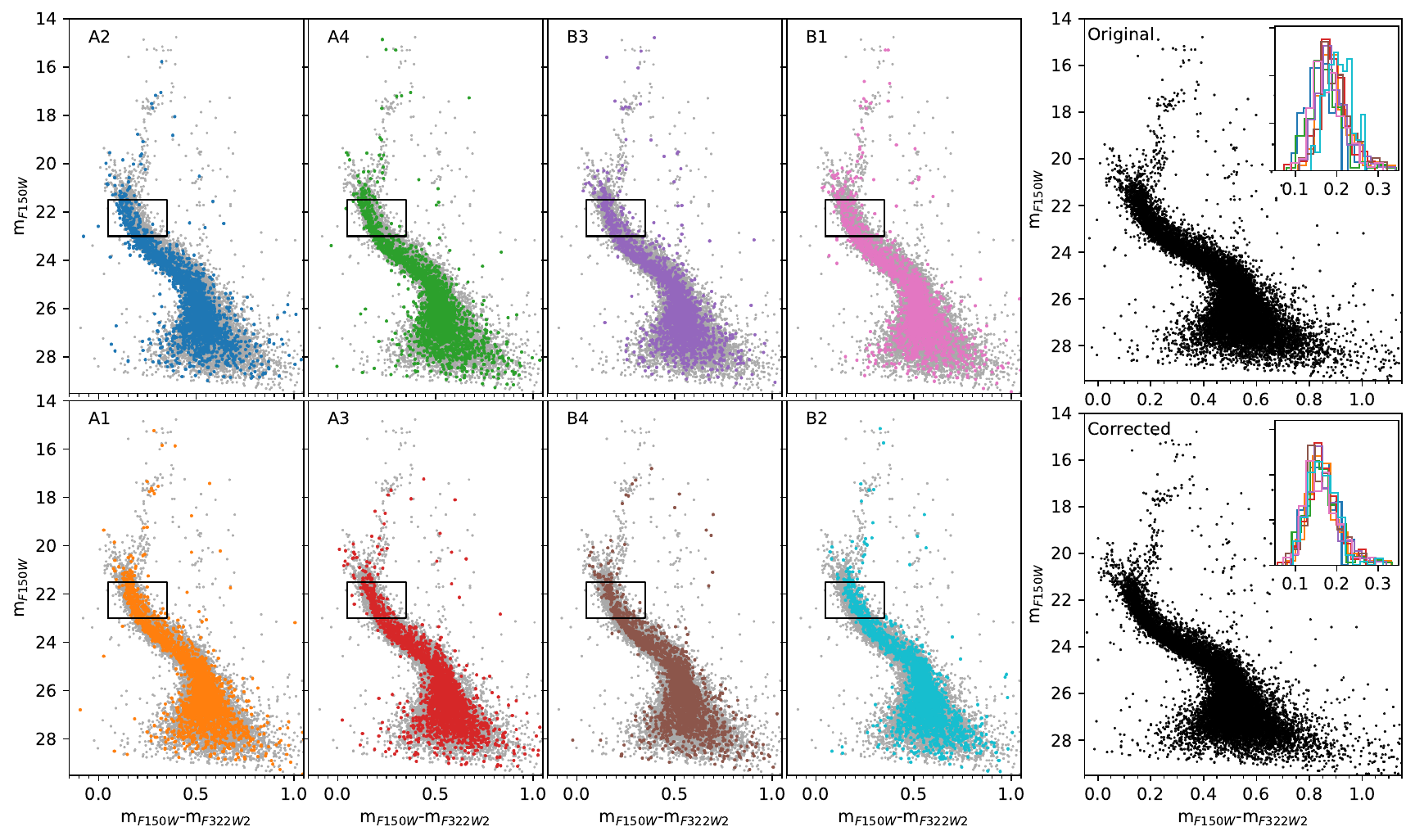}{0.99\textwidth}{}}
\vspace{-0.5cm}
\caption{\textbf{Left: }Per-detector CMDs illustrating the detector-dependent photometric offsets seen in the catalog output by DOLPHOT.  In each panel, stars in the labeled NIRCam SW chip are shown in color, overplotted on stars in the remaining chips.  The box illustrates the CMD region used to correct for these offsets by calculating the 3$\sigma$-clipped median color. \textbf{Right: }CMDs of all chips shown before (top) and after (bottom) the per-chip corrections were applied.  The insets show histograms of the color distribution in the CMD selection box (shown in black in the left panels), color-coded by SW detector.  The color distributions are shown before (top) and after (bottom) corrections.
\label{colorcorr_chip_fig}}
\end{figure}

The photometric catalogs output by DOLPHOT exhibited detector-dependent photometric offsets, displayed in Fig.~\ref{colorcorr_chip_fig}.  To correct for these offsets, we selected a relatively vertical region of the CMD (shown in black boxes in Fig.~\ref{colorcorr_chip_fig}) and calculated the 3-$\sigma$-clipped median color in each SW chip.  To correct for per-chip offsets in median color, we perform a two step process allowing for SW-detector-dependent offsets in F150W as well as LW-detector-dependent offsets in F322W2: For each LW module, we choose a reference chip (A2 and B1) and correct the remaining three chips to the median color of the reference chip by shifting their F150W magnitudes.  Next, we compare the corrected colors in each of the LW modules, and shift the F322W2 magnitudes of sources in module B (detector B5) to match those of module A (detector A5).  The per-detector offsets applied are listed in Table \ref{zpttab}, and CMDs displaying all sources passing our photometric quality cuts (see Sect.~\ref{photsect}) are shown before and after the correction in the right-hand panels of Fig.~\ref{colorcorr_chip_fig}.  To estimate the uncertainty on the applied per-detector corrections, we repeated the above procedure using an alternate CMD selection box located on the lower main sequence (25.5$\leq$F150W$\leq$27.0), which is also nearly vertical in the CMD, finding that the per-chip corrections differed by $\leq$0.006 mag in all cases with a median absolute deviation of 0.003 mag.

\begin{deluxetable}{cr}
\tablecaption{Detector-Dependent Corrections to Photometry\label{zpttab}}
\tablehead{
\colhead{SW Detectors} & \colhead{$\Delta$F150W (mag)}}
\startdata
A2-A1 & -0.026 \\
A2-A3 & -0.027 \\
A2-A4 & -0.010 \\
B1-B2 & -0.026 \\
B1-B3 & -0.004 \\
B1-B4 & -0.014 \\
\hline
LW Detectors & $\Delta$F322W2 (mag)\\
\hline
A5-B5 & 0.018
\enddata
\tablecomments{SW photometry is referred to detectors A2 and B1, and LW photometry is referred to detector A5 (see text for details).}
\end{deluxetable}

\section{Background Galaxy Catalogs \label{bgcatsect}}

\begin{figure}[b!]
\gridline{\fig{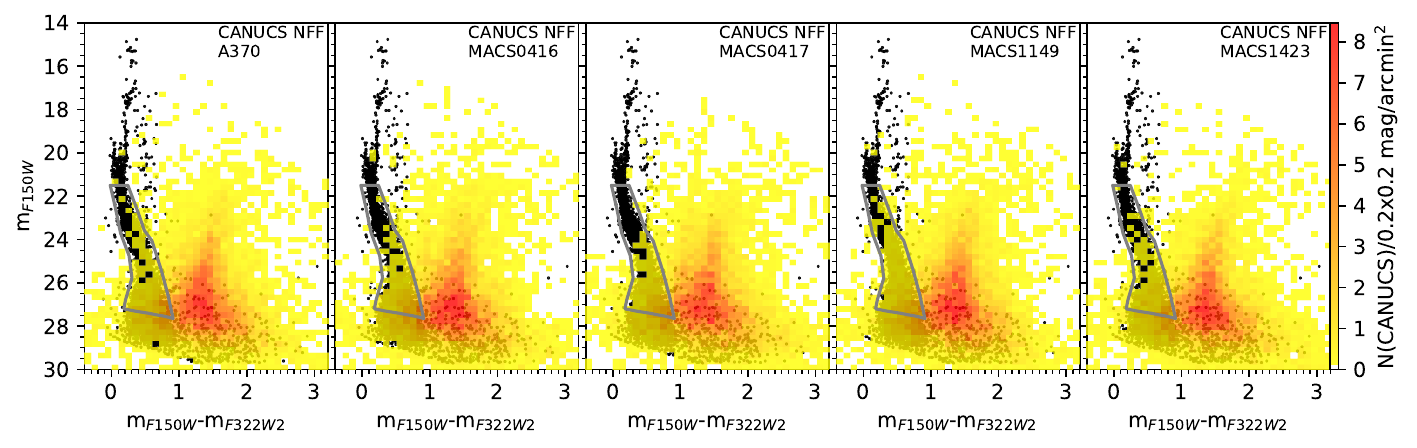}{0.99\textwidth}{}}
\vspace{-0.5cm}
\caption{Color-magnitide distribution of sources from each of the individual CANUCS NFFs.  Sources in our SMC photometric catalog passing our quality cuts are shown as black points, and the CMD region we use for our IMF analysis is indicated by a grey line.
\label{bggals_ind_fig}}
\end{figure}

The color-magnitude distribution of sources in each of the five individual CANUCS NIRCam flanking fields (NFFs) are shown in Fig.~\ref{bggals_ind_fig}, obtained from the publicly available catalogs.\footnote{\url{https://niriss.github.io}}  Because our concern is primarily with compact point-like sources masquerading as stars, we used fluxes calculated in the \texttt{COLOR03} aperture, intentionally imposing only minimal cuts on the catalogs to obtain as conservative a background galaxy selection as possible.  To this end, we retained all sources with valid fluxes in F150W, finding that none of them had \texttt{FLAG\_BCGCONTAM}=1, or \texttt{FLAG\_BPR\_MASTER} or \texttt{FLAG\_FOV\_MASTER}=1 in the relevant filters (flagging severe background contamination, pixels outside the field of view, and bad pixels respectively).  We also have not corrected the CANUCS fields for potential Galactic foreground contaminants towards each of the NFFs.  Although these may be flagged in the CANUCS catalogs by \texttt{FLAG\_POINTSRC} and/or \texttt{FLAG\_GAIA}, \citet{sarrouh25} clarify that this flagging was done only down to a flux of 100 nJy (m$_{\rm F150W}$$\sim$25.17) and may include true galaxies that appear point-source-like in their catalogs.  In any case, given the high Galactic latitudes of the CANUCS fields ($|$B$|$$>$39$^{\circ}$), their foreground contamination is predicted to be negligible ($<$0.2\% faintward of m$_{F150W}$=25) by the TRILEGAL galaxy model even if it underestimates the true projected density of foreground contaminants as in Sect.~\ref{fgsect}.

\section{Validating the Fitting Technique \label{validsect}}

\begin{figure}[h!]
\gridline{\fig{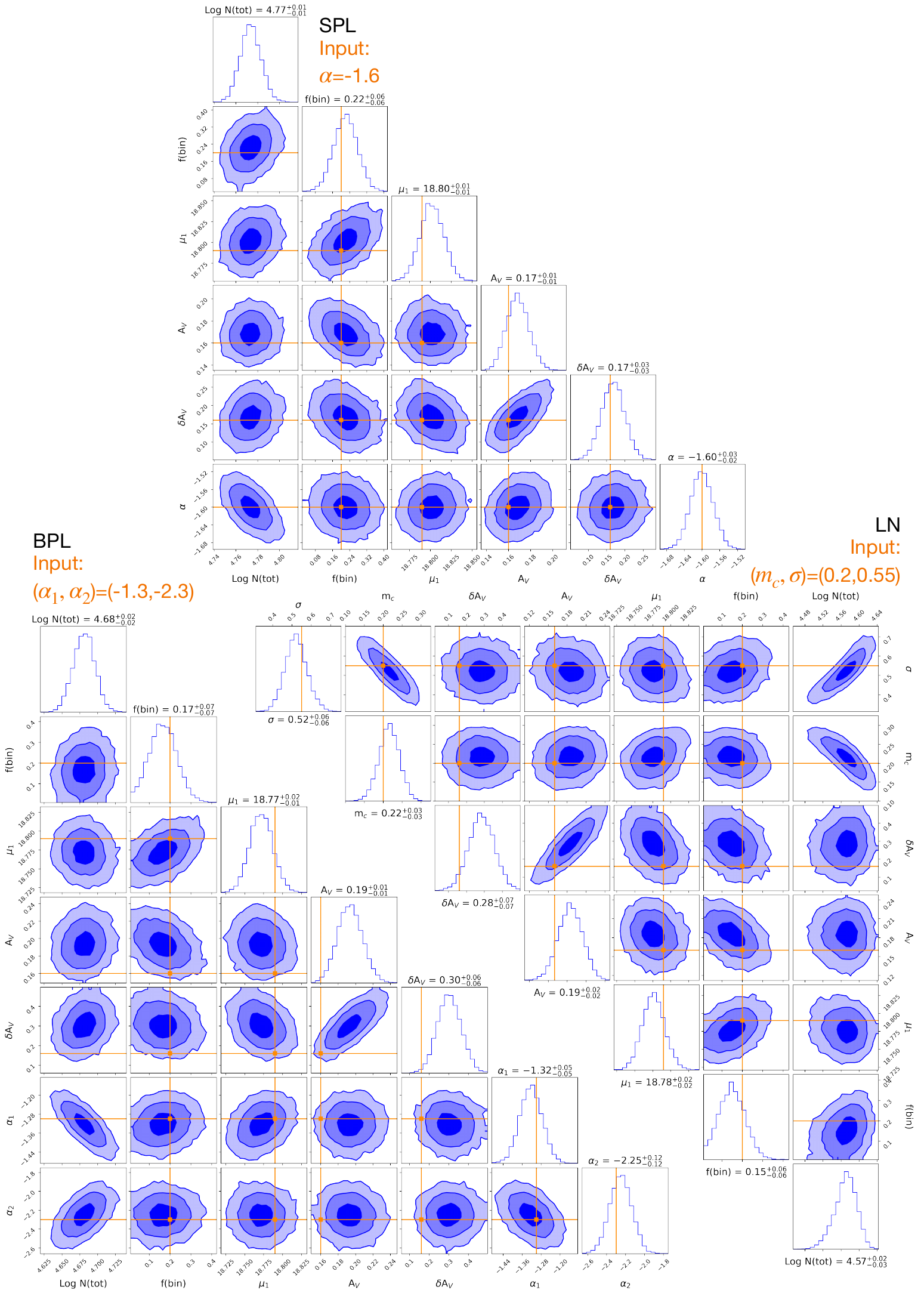}{0.842\textwidth}{}}
\vspace{-0.9cm}
\caption{Simulation results comparing the posterior distributions obtained (blue) to input parameters (orange) for the three different functional forms of the IMF we examine (SPL, BPL, IMF).  While the recovered distance and extinction vary from their input values by more than 1-$\sigma$ in some cases for the BPL and LN fits, the IMF parameters (and the binary fraction) are recovered to within 1-$\sigma$ in all cases.
\label{sim_fig}}
\end{figure}

To test our fitting technique described in Sect.~\ref{methodsect}, we applied this methodology to synthetic photometry generated using the assumptions in Sect.~\ref{inputsect} and known IMF input parameters.
These include a slope of $\alpha$=$-$1.6 for the SPL IMF, the \citet{kroupa01} values for the BPL IMF ($\alpha_{1}$=$-$1.3, $\alpha_{2}$=$-$2.3) and the \citet{chabrier05} values for the LN IMF ($m_{c}$=0.2, $\sigma$=0.55).  In addition, we assumed input values of f(bin)=0.2 based on our results in Sect.~\ref{resultsect}, the double Gaussian line-of-sight stellar distance distribution from \citet{elyoussoufi21}, which has $\mu_{1}$=18.79, and a lognormal extinction distribution with input values of (A$_{V}$, $\delta$A$_{V}$) = (0.16, 0.16) mag \citep{skowron21}.  We find that in all cases, the input IMF parameters and binary fraction were recovered to within their 68\% confidence intervals, although this was not always true of $\mu_{1}$, A$_{V}$ and $\delta$A$_{V}$ for the BPL and LN fits.  In Fig.~\ref{sim_fig}, we show the full output posterior distributions in blue for all free parameters corresponding to each of the three IMF forms analyzed (SPL, BPL, LN), compared with their input values shown using orange lines.  

\section{The Impact of Fitting Assumptions on IMF Parameters \label{testsect}}

Here we test the sensitivity of our IMF parameters to various input assumptions in Sect.~\ref{methodsect} producing the results presented in Sect.~\ref{resultsect}, which we refer to here as the \enquote{baseline} set of results.  We have varied the following assumptions, which are compared to our baseline set of IMF parameters in Fig.~\ref{test_fig}: \newline

\noindent \textit{CMD region used for IMF fitting:} To assess whether our results are systematically impacted by our choice of the 50\% photometric completeness limit to bound our chosen CMD region at the faint end, we reperformed our fits using the 20\% and 80\% completeness limits to set the faint end of our CMD region instead. \newline

\noindent \textit{Distance distribution:} To test the sensitivity of our results to the assumed line-of-sight distance distribution, we reperformed our fits using the double Gaussian parameters from \citet[][see their Table 3 and Fig.~B2]{omkumar21}, which predict a greater separation and asymmetry for the two line-of-sight components than our baseline values from \citet{elyoussoufi21}.  Specifically, they find ($\mu_{1}$, $\mu_{2}-\mu_{1}$, $\sigma_{1}$, $\sigma_{2}$, $A_{2}/A_{1}$) = (18.65 mag, 0.49 mag, 0.21 mag, 0.17 mag, 0.39). \newline

\begin{figure}[h!]
\gridline{\fig{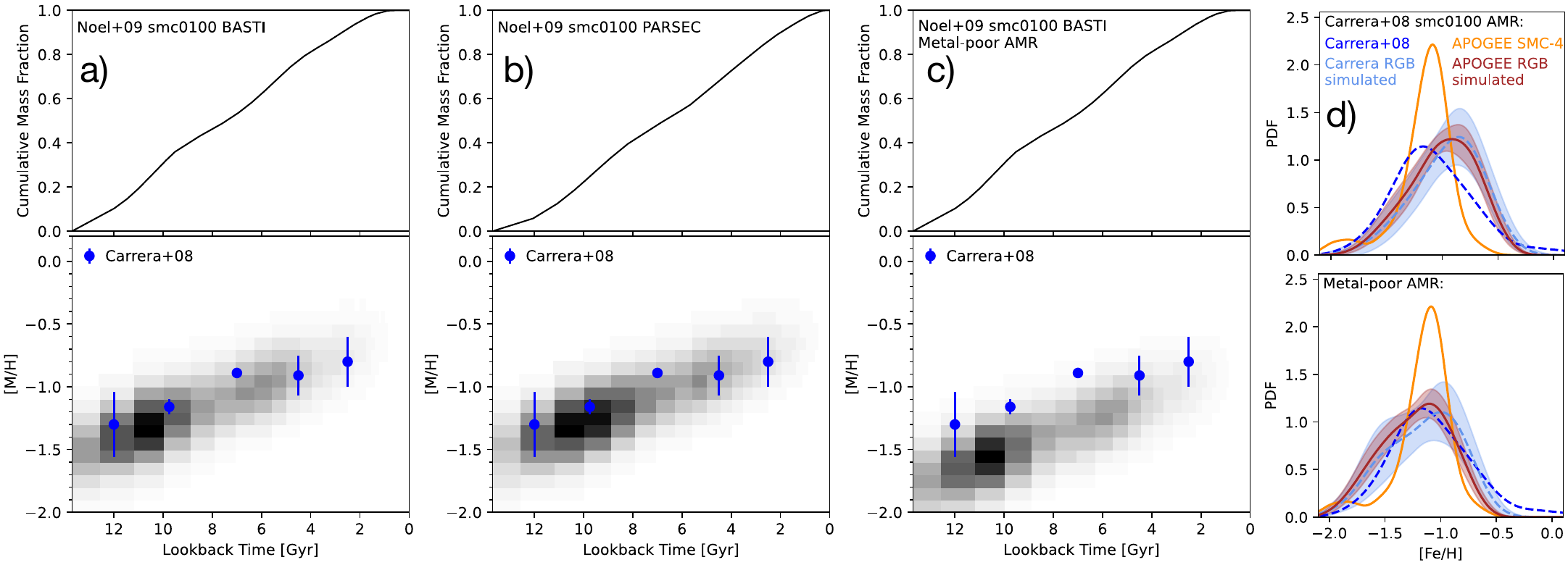}{0.99\textwidth}{}}
\vspace{-0.5cm}
\caption{Variations in the assumed SFH and AMR of our target field that were tested.  \textbf{(a)} The SFH (top) and AMR (bottom) assumed for our \enquote{baseline} analysis in Sects.~\ref{methodsect}-\ref{resultsect}, based on results for the co-spatial smc0100 field from \citet{carrera08} and \citet{noel09}.  \textbf{(b)} Same, but using the PARSEC-based SFH from \citet{noel09} rather than their BaSTI-based SFH.  \textbf{(c)} Same as (a), but assuming a more metal-poor AMR.  \textbf{(d)} In the top panel, we compare spectroscopic metallicity distributions from APOGEE DR17 and \citet{carrera08} (solid orange and dashed blue lines respectively) to the metallicity distributions obtained from synthetic photometry (solid brown and dashed cyan lines respectively), producing a metallicity distribution more metal-rich than observed.  In the lower panel, we perform the same comparison, but assuming the \enquote{metal-poor} AMR shown in panel (c), yielding metallicity distributions in good agreement with spectroscopy.
\label{sfhamr_fig}}
\end{figure}

\begin{figure}
\gridline{\fig{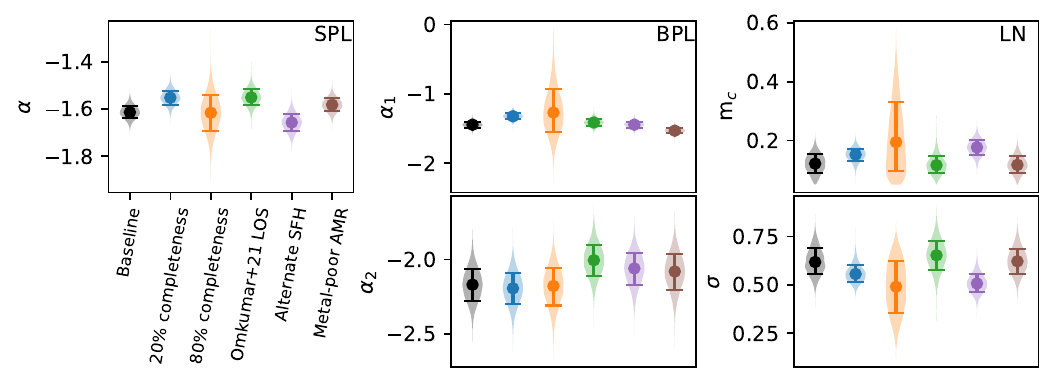}{0.99\textwidth}{}}
\caption{Comparison between the \enquote{baseline} values for the IMF parameters reported in Sect.~\ref{resultsect} versus those obtained for various test cases in which we vary our input assumptions for IMF fitting.  Results are color coded by test case, labeled under the first panel.  Filled circles and errorbars correspond to the median and 68\% confidence interval of each posterior distribution, indicated with shading.  The recovered IMF parameters are not sensitive to changes in our input assumptions at a statistically significant level.    
\label{test_fig}}
\end{figure}

\noindent \textit{Star formation history:}  For our analysis, we assumed the BaSTI-based SFH from \citet{noel09}, shown in the top panel of Fig.~\ref{sfhamr_fig}a.  To test the impact of reasonable variations in the assumed SFH, we reperformed our fits using the PARSEC-based (rather than BaSTI-based) SFH from \citet{noel09}, shown in Fig.~\ref{sfhamr_fig}b. \newline

\noindent \textit{Chemical Evolution History:} In the top panel of Fig.~\ref{sfhamr_fig}d, we show two spectroscopic metallicity distribution functions (MDFs) obtained from regions that are co-located with our target field.  Specifically, the blue dotted line indicates the MDF obtained for the smc0100 field by \citet{carrera08}, after transforming from the \citet{cg97} metallicity scale following \citet{c09}.  The orange solid line indicates the MDF obtained from the SMC-4 APOGEE field after applying membership criteria from \citet{nidever20}.  The two spectroscopic MDFs have (inverse-square-weighted) means of $\langle$[Fe/H]$\rangle$=$-$1.06$\pm$0.06 and $-$1.12$\pm$0.01 dex respectively, and median [Fe/H] of $-$1.14 and $-$1.11 dex respectively.  These spectroscopic MDFs are compared with the MDFs obtained from CMD-selected RGB stars generated over 1,000 realizations of synthetic photometry (i.e., assuming the AMR shown in the lower panel of Fig.~\ref{sfhamr_fig}d) in either the $I,(B-I)$ CMD used by \citet{carrera08}, shown using a light blue line, or the $(K_{S},J-K_{S})$ CMD, shown using a brown line.  In both cases, synthetic CMD realizations yield RGB MDFs more metal-rich than the observed spectroscopic metallicity distributions.  Therefore, we vary our assumed AMR, shifting the \enquote{baseline} AMR more metal-poor by 0.25 dex.  The resultant RGB MDFs from synthetic photometry assuming this more metal-poor AMR are again compared to the spectroscopic MDFs in the lower panel of Fig.~\ref{sfhamr_fig}d.  These synthetic MDFs, generated using the more metal-poor AMR, show better agreement with spectroscopic values, with median [Fe/H]=$-$1.15$\pm$0.16 and $-$1.20$\pm$0.07 using RGB stellar samples selected from $I,(B-I)$ or $K_{S},(J-K_{S})$ photometry respectively.  Although it is not immediately clear why the \citet{carrera08} AMR would produce an RGB MDF discrepant from their distribution of spectroscopic metallicities of individual RGB targets by $\sim$0.2-0.3 dex, possibilities include updates to the BaSTI evolutionary models used by \citet{carrera08} to obtain isochrone-based ages (their per-star ages were estimated using a polynomial relation with an uncertainty of 0.4 dex in Log age) as well as differences in assumed metallicity scales (although had we retained the \citealt{cg97} metallicity scale, the discrepancy would have worsened, not improved).  In any case, Fig.~\ref{test_fig} indicates that our assumption of the more metal-poor AMR (shown in the lower panel of Fig.~\ref{sfhamr_fig}c) has a statistically negligible impact on the IMF parameters.

More generally, while uncertainties on the IMF parameters increase in some cases when fitting shallower data (as predicted in, e.g., \citealt{elbadry17}), the changes in input assumptions we have tested do not result in statistically significant variations in recovered IMF parameters compared to the baseline values reported in Sect.~\ref{resultsect}, Table \ref{imfpartab} and Figs.~\ref{corner_fig}-\ref{resids_fig}.

\end{document}